\documentclass[aps,prl,reprint,superscriptaddress,floatfix]{revtex4-2}
\usepackage{graphicx}% Include figure files
\usepackage[colorlinks=true, linkcolor=blue, citecolor=blue, urlcolor=blue]{hyperref}
\usepackage{amsmath}

\begin{document}

\title{Intermittent chaos in an optomechanical resonator}

\author{Yue Huo}
\affiliation{Department of Automation, Tsinghua University, Beijing 100084, China}
\author{Zhe Wang}
\affiliation{Department of Automation, Tsinghua University, Beijing 100084, China}
\author{Zhenning Yang}
\affiliation{School of Automation Science and Engineering, Xi'an Jiaotong University, Xi'an 710049, China}
\affiliation{MOE Key Laboratory for Intelligent Networks and Network Security, Xi'an Jiaotong University, Xi'an 710049, China}
\author{Xiaohe Tang}
\affiliation{Department of Automation, Tsinghua University, Beijing 100084, China}
\author{Deng-Wei Zhang}
\affiliation{Department of Mathematics and Physics, Luoyang Institute of Science and Technology, Luoyang 471023, China}
\author{Qianchuan Zhao}
\affiliation{Department of Automation, Tsinghua University, Beijing 100084, China}
\author{Wenjie Wan}
\affiliation{University of Michigan-Shanghai Jiao Tong University Joint Institute, Shanghai Jiao Tong University, Shanghai 200240, China}
\affiliation{Department of Physics and Astronomy, Shanghai Jiao Tong University, Shanghai 200240, China}
\author{Yu-xi Liu}
\affiliation{School of Integrated Circuits, Tsinghua University, Beijing 100084, China}
\affiliation{Frontier Science Center for Quantum Information, Beijing 100084, China}
\author{Xin-You Lü}
\affiliation{School of Physics and Institute for Quantum Science and Engineering, Huazhong University of Science and Technology, Wuhan 430074, China}
\affiliation{Wuhan Institute of Quantum Technology, Wuhan 430074, China}
\author{Guangming Zhao}
\affiliation{State Key Laboratory of Optoelectronic Materials and Devices, Institute of Semiconductors, Chinese Academy of Sciences, Beijing, 100083, China}
\author{Liang Lu}
\affiliation{School of Physics and Optoelectronic Engineering, Anhui University, Hefei 230601, China}
\affiliation{National Key Laboratory of Opto-Electronic Information Acquisition and Protection Technology, Anhui University, Hefei 230601, China}
\author{Jing Zhang}
\email{zhangjing2022@xjtu.edu.cn}
\affiliation{School of Automation Science and Engineering, Xi'an Jiaotong University, Xi'an 710049, China}
\affiliation{MOE Key Laboratory for Intelligent Networks and Network Security, Xi'an Jiaotong University, Xi'an 710049, China}

\begin{abstract}
Chaos is a fundamental phenomenon in nonlinear dynamics, manifesting as irregular and unpredictable behavior across various physical systems. Among the diverse routes to chaos, intermittent chaos is a distinct transition pathway, characterized by the temporal or spatial alternation between periodic and chaotic motions. Here, we experimentally demonstrate, for the first time, optomechanically induced intermittent chaos in an optical whispering-gallery-mode microresonator. Specifically, the system evolves from stable periodic oscillation through an intermittent-chaos regime before fully developing into chaotic motion. As system parameters vary, the proportion of chaotic motion in the time-domain increases asymptotically until chaotic dynamics dominates entirely. Moreover, it is counterintuitive that, intermittent chaos can act as noise of a favorable intensity compared with purely periodic or fully chaotic states, and enhance rather than reduce system's responses in nonlinear ultrasonic detection. These findings not only deepen the comprehensive understanding of chaos formation but also broaden its potential applications in high-precision sensing and information processing.
\end{abstract}

\maketitle

\textit{Introduction}---Chaos is a special type of nonlinear dynamics that is highly sensitive to the initial conditions of a system, and exhibits uncertainty over time \cite{jiangChaosassistedBroadbandMomentum2017,podolskiyChaoticMicrolasersBased,jiangCoherentControlChaotic2024,xiaoTunnelinginducedTransparencyChaotic2013}. It has been used in various applications, such as secure communication \cite{hayesCommunicatingChaos1993,vanwiggerenCommunicationChaoticLasers1998}, random number generation \cite{uchidaFastPhysicalRandom2008}, and signal detection \cite{guanyuwangApplicationChaoticOscillators1999}. As the system parameters vary, chaotic dynamics typically emerges through an evolutionary process that begins with periodic motion and proceeds via period-doubling bifurcations, eventually leading to chaos \cite{carmonChaoticQuiveringMicronScaled2007,bakemeierRouteChaosOptomechanics2015,monifiOptomechanicallyInducedStochastic2016a}. Besides the period-doubling bifurcation route, intermittent chaos represents another distinctive route to chaos, and features the alternating appearance of periodic and chaotic motions in temporal or spatial evolution. Intermittent chaos was first discovered in the Lorenz model \cite{mannevilleIntermittencyLorenzModel1979}, in which the intermittent transition from a limit cycle to a strange attractor was demonstrated. As a unique nonlinear phenomenon, intermittent chaos has been observed in different physical systems \cite{pomeauIntermittentTransitionTurbulencea,qinChaosBifurcationsPeriodic1989,fu-shengIntermittentChaosPulsating,cherninIntermittentChaosThreebody1998,hramovIntermittencyRouteChaos2015,suzukiPeriodicQuasiperiodicChaotic2016,bubanjaIntermittentChaosBray2016,aizawaSymbolicDynamicsApproach1983,bruntonChaosIntermittentlyForced2017,zhangIntermittentChaosCavity2020}, including fluid \cite{benziIntermittencyCoherentStructures1986}, laser \cite{tangLaserDynamicsTypeI1992}, electric circuit \cite{shirahamaIntermittentChaosMutually1998}, and artificial neuron \cite{hoppensteadtIntermittentChaosSelforganization1989,wojtusiakIntermittentMetastableChaos2021}. However, its practical application has so far been mainly limited to random number generation \cite{boscoRandomNumberGeneration2017}. Exploring the underlying mechanisms and distinctive characteristics of this route deepens our understanding of the nature of chaos, and offers new perspectives for its potential applications.

Whispering-gallery-mode (WGM) microcavity has recently attracted wide attention due to its high optical quality (Q)-factor and small mode volume \cite{armaniUltrahighQToroidMicrocavity2003}, and been employed in various applications, such as sensing \cite{chenExceptionalPointsEnhance2017,zhuOnchipSingleNanoparticle2010a,goldwinFastCavityenhancedAtom2011,suLabelfreeDetectionSingle2016,liQuantumEnhancedOptomechanical2018,forstnerCavityOptomechanicalMagnetometer2012,basiri-esfahaniPrecisionUltrasoundSensing2019,xuWirelessWhisperinggallerymodeSensor2018,xuPhonesizedWhisperinggalleryMicroresonator2016,liaoOpticalWhisperinggalleryMode2021,rosenblumCavityRingupSpectroscopy2015,laiObservationExceptionalpointenhancedSagnac2019a,laiEarthRotationMeasured2020a,qvarfortGravimetryNonlinearOptomechanics2018}, high-speed communication \cite{marin-palomoMicroresonatorbasedSolitonsMassively2017,spencerOpticalfrequencySynthesizerUsing2018}, and parallel computation \cite{feldmannParallelConvolutionalProcessing2021a}. Driven by the optical radiation pressure \cite{aspelmeyerCavityOptomechanics2014}, the intrinsic mechanical mode of the microcavity can be excited by the optical mode, and the enhanced light–matter interactions demonstrate abundant nonlinear optomechanical effects, such as phonon laser \cite{grudininPhononLaserAction2010,zhangPhononLaserOperating2018a}, optomechanical soliton \cite{zhangOptomechanicalDissipativeSolitons2021}, and optomechanical chaos \cite{carmonTemporalBehaviorRadiationPressureInduced2005,carmonChaoticQuiveringMicronScaled2007,bakemeierRouteChaosOptomechanics2015,monifiOptomechanicallyInducedStochastic2016a,luSymmetryBreakingChaosOptomechanics2015}. Optomechanical chaos was first experimentally observed in an ultra-high-Q toroidal microcavity \cite{carmonTemporalBehaviorRadiationPressureInduced2005}, and was later both theoretically analyzed \cite{bakemeierRouteChaosOptomechanics2015} and experimentally demonstrated \cite{carmonChaoticQuiveringMicronScaled2007,monifiOptomechanicallyInducedStochastic2016a} following the period-doubling bifurcation route to chaos in cavity optomechanics. Furthermore, optomechanical chaos induced by the parity-time (PT) symmetry breaking was investigated in coupled cavities, which features a controllable PT-symmetry phase transition \cite{luSymmetryBreakingChaosOptomechanics2015}. However, the mechanism of intermittent chaos and its transition route in optomechanical system is yet to be studied. Benefiting from its multi-mode nature and flexible parameter tunability, the optomechanical system offers favorable conditions for the observation of this phenomenon.

In this letter, we experimentally demonstrate, for the first time, optomechanically induced intermittent chaos in a WGM microtoroidal resonator. By tracking the route to chaos and analyzing system dynamics in both time and frequency domains, we reveal that intermittent chaos features the alternating appearance of periodic and chaotic motions over time. Depending on the system parameters, the dynamics can be divided into three regimes: periodic, intermittent-chaos and chaotic regimes. As the frequency detuning between the cavity optical mode and input field increases, chaotic dynamics begin to emerge within the periodic motion. Further increase in frequency detuning leads to an increased proportion of time spent in the chaotic motion, eventually resulting in a fully chaotic state. In the intermittent-chaos regime, the spectral base beneath the isolated peaks in the periodic regime becomes elevated and develops a broadened hollow-like structure, reflecting the coexistence of periodic and chaotic dynamics. Counterintuitively, intermittent chaos can act as noise of a certain intensity, and is leveraged to detect the cavity’s responses to ultrasonic, resulting in significant improvements in both signal-to-noise ratio (SNR) and noise equivalent power (NEP).

\begin{figure}
	\includegraphics[width=0.98\linewidth]{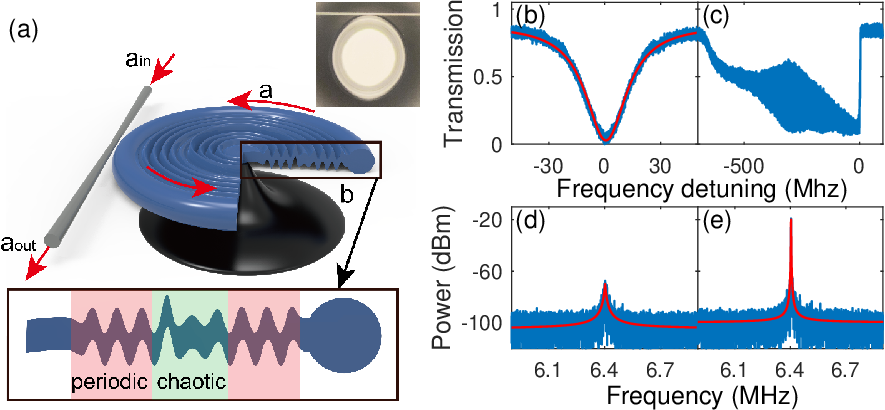}
	\caption{\label{fig:one}Optomechanics-induced intermittent chaos. (a) The cross section of the microtoroid showing the intermittent-chaos vibration. The inset shows the top view of a toroidal microcavity. (b) Optical transmission spectrum obtained by scanning the wavelength of a tunable laser with a power below the mechanical oscillation threshold. (c) Optical transmission spectrum above the threshold. Thermally induced fluctuations due to the mechanical oscillations emerge. (d) Spectrum of the mechanical mode below the threshold. (e) Spectrum of the mechanical mode above the threshold.}
\end{figure}

\textit{Intermittent chaos}---A WGM optomechanical resonator supports both an optical mode $a$ and a mechanical mode $b$, as illustrated in Fig.~\ref{fig:one}(a). The input laser $a_{in}$ is coupled into the microcavity via the evanescent field of a tapered fiber and circulates along the cavity’s periphery through total internal reflection. When the input optical power exceeds a certain threshold, radiation pressure from the optical field excites mechanical vibrations in the microcavity. In turn, the mechanical deformation modifies the effective cavity length and alters the optical resonance frequency. This mutual coupling gives rise to nonlinear light-matter interactions, and induces optomechanical intermittent chaos. The cross section of the microcavity in Fig.~\ref{fig:one}(a) shows intermittent-chaos vibration of the microtoroidal optomechanical resonator. Intermittent chaos features alternating appearance of periodic and chaotic motions, when the acoustic wave propagates along the surface of the microtoroidal resonator. A similar interplay between order and disorder occurs in hydrodynamics, where fluid flow alternates between laminar (periodic) and turbulent (chaotic) states. This analogy highlights the universality of intermittent chaos across physical systems. The transmission spectrum of the optical mode with quality factor of $1.07\times10^7$ is shown in Fig.~\ref{fig:one}(b). When the power of the input laser is increased, the mechanical mode is motivated, and the transmission spectrum with thermal effect is also shown in Fig.~\ref{fig:one}(c). The spectrum of the mechanical mode with frequency of 21.5 MHz and quality factor of $2.3\times10^3$ is shown in Fig.~\ref{fig:one}(d), when the laser power is below the
threshold. For power below the threshold, the linewidth is significantly narrowed as shown in Fig.~\ref{fig:one}(e).

\begin{figure}
	\includegraphics[width=0.98\linewidth]{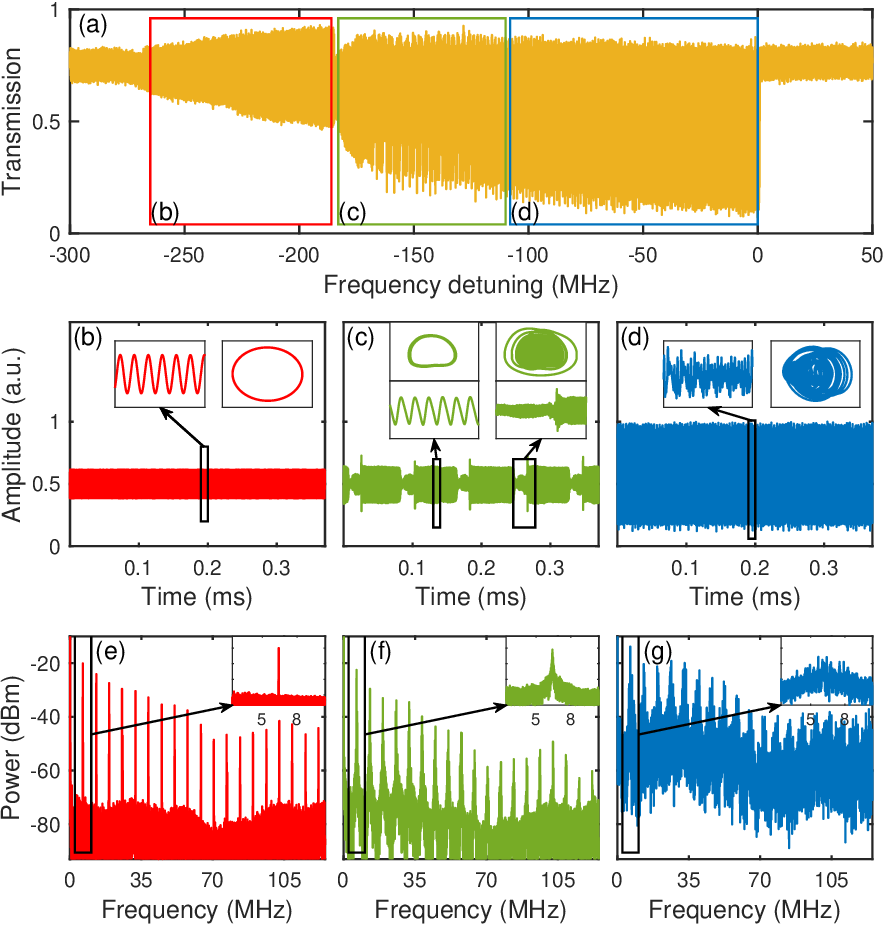}
	\caption{\label{fig:two}Transition from periodic regime to chaotic regime mediated by intermittent chaos. (a) Different regimes as a function of frequency detuning between the cavity optical field and the input field. With the increase of frequency detuning, the output field transits from the periodic regime, then is mediated by intermittent-chaos regime, and finally gets into the chaotic regime. (b-d) Time-domain signals of the output fields in periodic (b), intermittent-chaos (c) and chaotic (d) regimes. The insets are the details of time evolution and the corresponding phase diagrams. (e-g) Spectra of output fields in periodic (e), intermittent-chaos (f) and chaotic (g) regimes. The insets show the details of the main peak of the mechanical mode.}
\end{figure}

By varying the frequency detuning, the optomechanical microresonator transits between three distinct regimes: periodic, intermittent-chaos and chaotic regimes, as illustrated in Fig.~\ref{fig:two}(a). In the periodic regime, the system exhibits a regular and repeating time-domain trajectory, as shown in Fig.~\ref{fig:two}(b), and the insets plot the details of time evolution and the corresponding phase diagram, revealing a single closed-loop orbit in the phase space. The spectrum of the output field exhibits isolated spectral lines, and the spectral base remains low at approximately –85 dBm, as shown in Fig.~\ref{fig:two}(e). As the frequency detuning increases, the system enters the intermittent-chaos regime, where the time-domain signal alternates between periodic and chaotic motions, as clearly illustrated in Fig.~\ref{fig:two}(c) and its insets. The trajectory of the chaotic part in the phase diagram spreads from a closed orbit to the surrounding area, in contrast to that of periodic part. The spectral base beneath the isolated spectral lines becomes elevated around the positions corresponding to those in the periodic regime, due to the influence of the chaotic part. Moreover, the elevated spectral base develops a hollow-like structure as presented in Fig.~\ref{fig:two}(f), and consequently, the spectral lines also become broadened, reflecting the coexistence of periodic and chaotic dynamics. As the system further evolves into the chaotic regime, the time-domain trajectory becomes completely aperiodic and exhibits random variations, as shown in Fig.~\ref{fig:two}(d), and the phase diagram in the inset becomes densely filled. The entire spectral base becomes elevated, and the influence of the original periodic peaks becomes almost indiscernible, as shown in Fig.~\ref{fig:two}(g). 

To quantify the alternating appearance of periodic and chaotic motions over time in the intermittent-chaos regime, we employ a periodic square wave as a reference, and these intermittent-chaos signals can be separated into two distinct segments in the time domain---periodic and chaotic parts. The time-domain expression of intermittent-chaos signals $x_I(t)$ can be approximated as
\begin{equation}
x_{I}(t)=x_{PS}(t)+x_{CS}(t)=x_{P}(t)\times s(t)+x_{C}(t)\times[1-s(t)],
\label{eq1}
\end{equation}
where $x_{PS}(t)=x_{P}(t)\times s(t)$ denotes the periodic part, and $x_{CS}(t)=x_{C}(t)\times[1-s(t)]$ denotes the chaotic part. Here, $x_{P}(t)$, $x_{C}(t)$, and $s(t)$ respectively represent the periodic signal, the chaotic signal, and the periodic-square-wave function, which is defined as
\begin{equation}
s(t) =
\begin{cases}
1, & \left|t\right| < T_0/2 \\
0, & T_0/2 < \left|t\right| \leq T_s/2.
\end{cases}
\label{eq2}
\end{equation}
$D=T_{0}/T_{s}$ in Eq.~(\ref{eq2}) is the duty cycle, where $T_{s}$ is the cycle period of the square wave and $T_{0}$ is the duration of chaotic part within one cycle. The duty cycle $D$ thus corresponds to the proportion of time spent in chaotic motion. As the duty cycle $D$ increases, the system will gradually transit from a periodic state to a fully chaotic state via the intermittent-chaos regime.
\begin{figure}
	\includegraphics[width=0.98\linewidth]{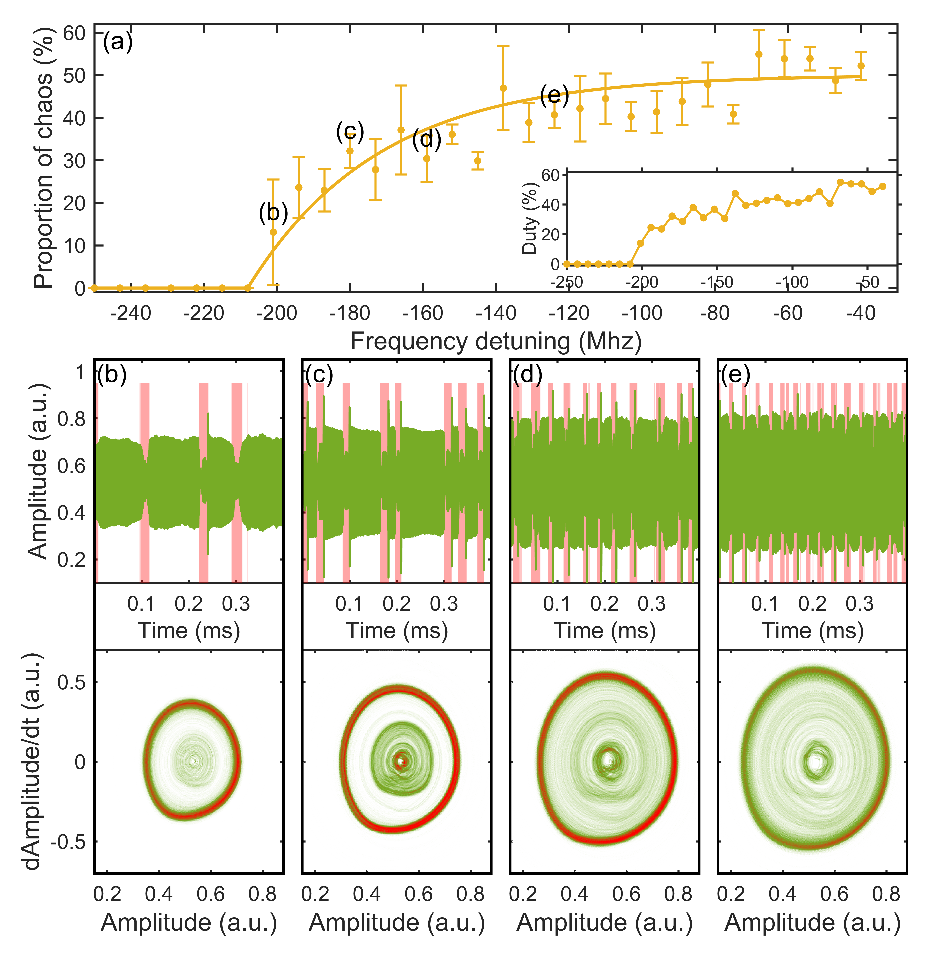}
	\caption{\label{fig:three}Mechanism of the route to chaos via intermittent chaos. (a) The experimental results of the proportion of time spent in the chaotic motion varying with frequency detuning. The inset shows the duty cycle variation of the periodic square wave to separate intermittent-chaos signal. (b-e) Time-domain signals (upper part) and the corresponding trajectory density distributions in the phase space (lower part) under different proportions of chaos. The chaotic parts in the time-domain signals are marked by a pink shadow. As the proportion of chaos increases, more and more trajectories in the phase space spreads from the closed orbit (colored in red) to the surrounding area, indicating a decreasing density of periodic trajectory and an increasing dominance of chaotic motion.}
\end{figure}

To experimentally explore the increasing trend of the chaotic proportion along the route to chaos through intermittent chaos, we continuously vary the frequency detuning while recording and analyzing the changes of the system states. As the frequency detuning increases, chaotic motion emerges, and its proportion initially grows rapidly from zero before gradually approaching saturation due to the nonlinear optomechanical interaction, as illustrated in Fig.~\ref{fig:three}(a). This nonlinear increase eventually drives the system into a fully chaotic state. The duty cycle $D$ of the periodic square wave used to separate the intermittent-chaos signal follows the same increasing trend as that of the chaotic proportion. During the transition from a periodic state to a chaotic state via the intermittent-chaos regime, the chaotic motions appear more and more frequently in the time-domain signal, as shown in the upper parts of the Fig.~\ref{fig:three}(b)-(e). Simultaneously, more and more trajectories in the phase space spread from the closed orbit to the surrounding region, indicating a decreasing proportion of periodic motion and an increasing dominance of chaotic motion, as shown in the lower parts of the Fig.~\ref{fig:three}(b)-(e). These experimental results clearly demonstrate the route to chaos through intermittent chaos, as theoretically discussed in Ref. \cite{zhangIntermittentChaosCavity2020}, which differs from the bifurcation route reported in Refs. \cite{carmonChaoticQuiveringMicronScaled2007,bakemeierRouteChaosOptomechanics2015,monifiOptomechanicallyInducedStochastic2016a}. During this process, periodic and chaotic motions coexist, with chaotic motion gradually becoming dominant. The quantitative analysis of the chaotic proportion reveals that, in contrast to the discrete period-doubling bifurcation, the intermittent route to chaos exhibits a continuous increase in chaotic motion.

\begin{figure}
	\includegraphics[width=0.98\linewidth]{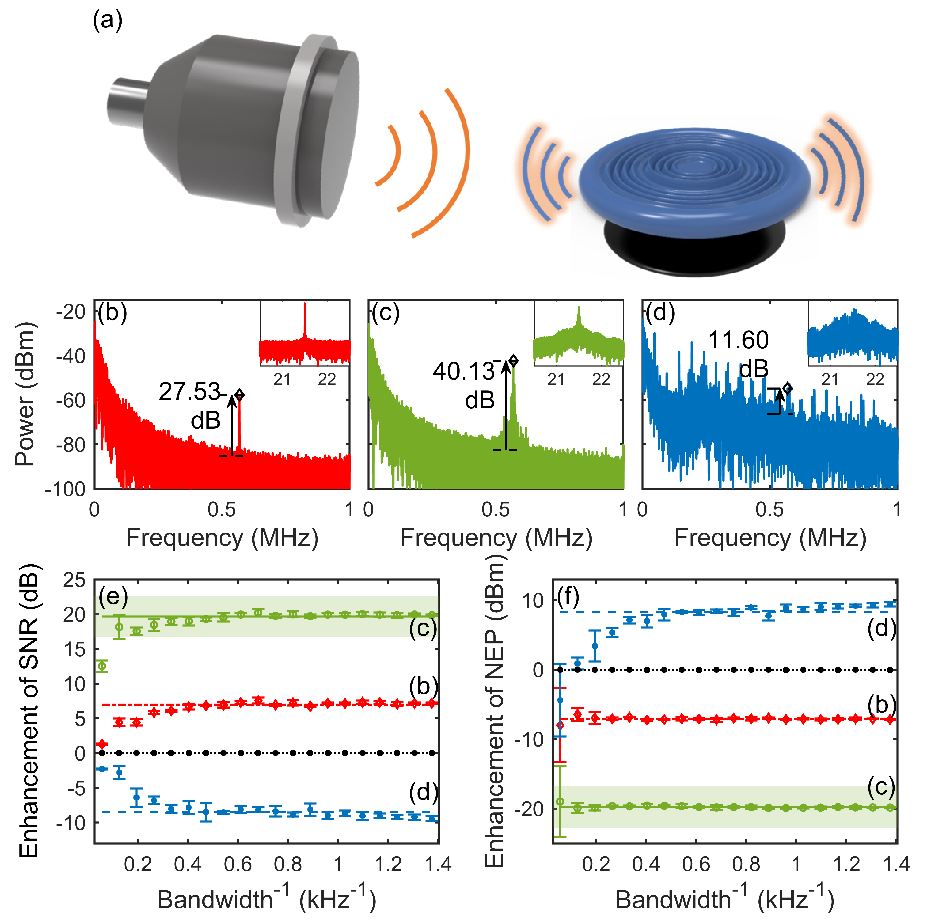}
	\caption{\label{fig:four}Ultrasonic sensing by the intermittent chaos. (a) Schematic diagram of the experimental set-up. The mechanical mode of the cavity has a frequency of 21.5 MHz, while the center frequency of the broadband ultrasonic signal is 570 kHz. (b-d) Spectra measuring the ultrasonic signal when the optomechanical system is in the periodic state (b), intermittent-chaos state (c) and chaotic state (d). The insets show the spectra of the cavity mechanical mode without the imposed ultrasonic signal. (e) The enhancement of the SNRs of ultrasonic response signals with different bandwidths, compared with the responses of a commercial ultrasonic probe. (f) The enhancement of the NEPs of ultrasonic response signals with different bandwidths, compared with the responses of a commercial ultrasonic probe. The black dotted line represents the baseline corresponding to the commercial ultrasound probe. The green solid line, the red dash-dotted line and the blue dashed line represent the responses in the intermittent-chaos, periodic and chaotic state.}
\end{figure}

\textit{Sensing}---Chaos, with its inherent randomness and broadband spectral characteristics, can be regarded as a source of noise, where stronger chaos corresponds to higher noise intensity. Conventionally, noise is considered disruptive to signal measurement. However, in nonlinear systems, noise of a certain intensity can paradoxically enhance signal detection capability and improve response sensitivity. This counterintuitive effect has been demonstrated in studies of stochastic resonance \cite{badzeyCoherentSignalAmplification2005,monifiOptomechanicallyInducedStochastic2016a} and coherence resonance \cite{liuCoherenceResonanceCoupled2001}, and further exploited to reduce the task complexity \cite{liPositiveIncentiveNoise2024}, improve synchronization stability \cite{shiNoiseenhancedStabilitySynchronized2025}, and enhance sensing response \cite{liStochasticExceptionalPoints2023}. Intermittent chaos, situated between the periodic and chaotic regimes, naturally acts as noise of a favorable intensity. When utilized for detecting an optomechanical cavity’s response to external signals, such as ultrasonic, shown in Fig.~\ref{fig:four}(a), this intermittent-chaos state enables significant improvements in sensing performance, outperforming the purely periodic and fully chaotic regimes.

When an external ultrasonic signal is imposed on the optomechanical microcavity, the cavity generates a response signal at the same frequency, which can be observed in the output spectrum. In the periodic state, the response peak induced by the ultrasonic is visible but remains relatively weak, with a limited signal-to-noise ratio (SNR), as shown in Fig.~\ref{fig:four}(b). When the system enters the intermittent-chaos state, the amplitude and SNR of the response signal are maximized, as shown in Fig.~\ref{fig:four}(c). This is attributable to the fact that intermittent chaos naturally introduces effective noise of a favorable intensity, which enhances the system responses and improves noise immunity. However, once the system evolves into the fully chaotic state, the entire spectral baseline is raised, and the response peak becomes buried in the noisy background,  as shown in Fig.~\ref{fig:four}(d). Consequently, the SNR deteriorates and the signal becomes difficult to discern. The SNR quantifies the strength of the response signal relative to background noise, and further serves as the basis for calculating the noise equivalent power (NEP). The NEP corresponds to the minimum ultrasonic power required to induce the response signal to rise above the noise floor, i.e., the condition where $\mathrm{SNR}=1$. A lower NEP indicates higher sensor sensitivity. The NEP is defined as
\begin{equation}
P_{\mathrm{NEP}}=\frac{P_u}{\mathrm{SNR}\times\sqrt{B}},
\label{eq3}
\end{equation}
where $P_u$ is the power of the imposed ultrasonic signal, and $B$ is the resolution bandwidth of the spectrum analyzer. Since the SNR is maximized in the intermittent-chaos state, the corresponding NEP is minimized, enabling the detection of much weaker signals. By contrast, the periodic state demands stronger input signal, while the chaotic state’s elevated noise floor increases the NEP and degrades performance. As shown in Fig.~\ref{fig:four}(e) and (f), the enhancements in both SNRs and NEPs, compared with the responses of a commercial ultrasonic probe, are most pronounced in the intermittent-chaos state.

\textit{Conclusion}---We have reported the first experimental demonstration of a distinct route to chaos in a WGM microtoroidal optomechanical resonator, where periodic motion transits into chaotic motion through an intermediate intermittent-chaos regime. This transition pathway fundamentally differs from the classical period-doubling bifurcation route, and is characterized by a continuous increase in the proportion of chaotic motion in time-domain signal. Beyond its fundamental significance, optomechanically induced intermittent chaos provides practical advantages for sensing applications. Acting as noise of a favorable intensity, it enhances both the SNR and NEP of the ultrasonic responses, outperforming the purely periodic and fully chaotic states, as well as a commercial ultrasonic probe. Overall, these findings contribute to a deeper understanding of chaos dynamics, and expand the application scopes of optomechanical systems in precision measurement and information transduction between optical and radio frequencies.

\textit{Acknowledgments}---J. Z. is supported by the National Natural Science Foundation of China (Grant No. 62433015), the Leading Scholar of Xi'an Jiaotong University, the Innovative Leading Talent Project of "Shuangqian plan" in Jiangxi Province, and the Tianfu Emei Plan-Provincial-University-Enterprise Cooperation Talent Special Project Funding. L. L. is supported by the National Natural Science Foundation of China (Grant No. 62275001). X.-Y. L. is supported by the National Science Fund for Distinguished Young Scholars of China (Grant No. 12425502), the National Key Research and Development Program of China (Grant No. 2021YFA1400700), and the Innovation Program for Quantum Science and Technology (Grant No. 2024ZD0301000).

\bibliography{refs}

\providecommand{\noopsort}[1]{}\providecommand{\singleletter}[1]{#1}%
\begin{thebibliography}{57}%
\makeatletter
\providecommand \@ifxundefined [1]{%
 \@ifx{#1\undefined}
}%
\providecommand \@ifnum [1]{%
 \ifnum #1\expandafter \@firstoftwo
 \else \expandafter \@secondoftwo
 \fi
}%
\providecommand \@ifx [1]{%
 \ifx #1\expandafter \@firstoftwo
 \else \expandafter \@secondoftwo
 \fi
}%
\providecommand \natexlab [1]{#1}%
\providecommand \enquote  [1]{``#1''}%
\providecommand \bibnamefont  [1]{#1}%
\providecommand \bibfnamefont [1]{#1}%
\providecommand \citenamefont [1]{#1}%
\providecommand \href@noop [0]{\@secondoftwo}%
\providecommand \href [0]{\begingroup \@sanitize@url \@href}%
\providecommand \@href[1]{\@@startlink{#1}\@@href}%
\providecommand \@@href[1]{\endgroup#1\@@endlink}%
\providecommand \@sanitize@url [0]{\catcode `\\12\catcode `\$12\catcode
  `\&12\catcode `\#12\catcode `\^12\catcode `\_12\catcode `\%12\relax}%
\providecommand \@@startlink[1]{}%
\providecommand \@@endlink[0]{}%
\providecommand \url  [0]{\begingroup\@sanitize@url \@url }%
\providecommand \@url [1]{\endgroup\@href {#1}{\urlprefix }}%
\providecommand \urlprefix  [0]{URL }%
\providecommand \Eprint [0]{\href }%
\providecommand \doibase [0]{https://doi.org/}%
\providecommand \selectlanguage [0]{\@gobble}%
\providecommand \bibinfo  [0]{\@secondoftwo}%
\providecommand \bibfield  [0]{\@secondoftwo}%
\providecommand \translation [1]{[#1]}%
\providecommand \BibitemOpen [0]{}%
\providecommand \bibitemStop [0]{}%
\providecommand \bibitemNoStop [0]{.\EOS\space}%
\providecommand \EOS [0]{\spacefactor3000\relax}%
\providecommand \BibitemShut  [1]{\csname bibitem#1\endcsname}%
\let\auto@bib@innerbib\@empty
%</preamble>
\bibitem [{\citenamefont {Jiang}\ \emph {et~al.}(2017)\citenamefont {Jiang},
  \citenamefont {Shao}, \citenamefont {Zhang}, \citenamefont {Yi},
  \citenamefont {Wiersig}, \citenamefont {Wang}, \citenamefont {Gong},
  \citenamefont {Lon{\v c}ar}, \citenamefont {Yang},\ and\ \citenamefont
  {Xiao}}]{jiangChaosassistedBroadbandMomentum2017}%
  \BibitemOpen
  \bibfield  {author} {\bibinfo {author} {\bibfnamefont {X.}~\bibnamefont
  {Jiang}}, \bibinfo {author} {\bibfnamefont {L.}~\bibnamefont {Shao}},
  \bibinfo {author} {\bibfnamefont {S.-X.}\ \bibnamefont {Zhang}}, \bibinfo
  {author} {\bibfnamefont {X.}~\bibnamefont {Yi}}, \bibinfo {author}
  {\bibfnamefont {J.}~\bibnamefont {Wiersig}}, \bibinfo {author} {\bibfnamefont
  {L.}~\bibnamefont {Wang}}, \bibinfo {author} {\bibfnamefont {Q.}~\bibnamefont
  {Gong}}, \bibinfo {author} {\bibfnamefont {M.}~\bibnamefont {Lon{\v c}ar}},
  \bibinfo {author} {\bibfnamefont {L.}~\bibnamefont {Yang}},\ and\ \bibinfo
  {author} {\bibfnamefont {Y.-F.}\ \bibnamefont {Xiao}},\ }\bibfield  {title}
  {\bibinfo {title} {Chaos-assisted broadband momentum transformation in
  optical microresonators},\ }\href {https://doi.org/10.1126/science.aao0763}
  {\bibfield  {journal} {\bibinfo  {journal} {Science}\ }\textbf {\bibinfo
  {volume} {358}},\ \bibinfo {pages} {344} (\bibinfo {year}
  {2017})}\BibitemShut {NoStop}%
\bibitem [{\citenamefont {Podolskiy}\ \emph {et~al.}(2004)\citenamefont
  {Podolskiy}, \citenamefont {Narimanov}, \citenamefont {Fang},\ and\
  \citenamefont {Cao}}]{podolskiyChaoticMicrolasersBased}%
  \BibitemOpen
  \bibfield  {author} {\bibinfo {author} {\bibfnamefont {V.~A.}\ \bibnamefont
  {Podolskiy}}, \bibinfo {author} {\bibfnamefont {E.}~\bibnamefont
  {Narimanov}}, \bibinfo {author} {\bibfnamefont {W.}~\bibnamefont {Fang}},\
  and\ \bibinfo {author} {\bibfnamefont {H.}~\bibnamefont {Cao}},\ }\bibfield
  {title} {\bibinfo {title} {Chaotic microlasers based on dynamical
  localization},\ }\href {https://doi.org/10.1073/pnas.0402805101} {\bibfield
  {journal} {\bibinfo  {journal} {Proc. Natl. Acad. Sci. U.S.A.}\ }\textbf
  {\bibinfo {volume} {101}},\ \bibinfo {pages} {10498} (\bibinfo {year}
  {2004})}\BibitemShut {NoStop}%
\bibitem [{\citenamefont {Jiang}\ \emph {et~al.}(2024)\citenamefont {Jiang},
  \citenamefont {Yin}, \citenamefont {Li}, \citenamefont {Quan}, \citenamefont
  {Goh}, \citenamefont {Cotrufo}, \citenamefont {Kullig}, \citenamefont
  {Wiersig},\ and\ \citenamefont {Al{\`u}}}]{jiangCoherentControlChaotic2024}%
  \BibitemOpen
  \bibfield  {author} {\bibinfo {author} {\bibfnamefont {X.}~\bibnamefont
  {Jiang}}, \bibinfo {author} {\bibfnamefont {S.}~\bibnamefont {Yin}}, \bibinfo
  {author} {\bibfnamefont {H.}~\bibnamefont {Li}}, \bibinfo {author}
  {\bibfnamefont {J.}~\bibnamefont {Quan}}, \bibinfo {author} {\bibfnamefont
  {H.}~\bibnamefont {Goh}}, \bibinfo {author} {\bibfnamefont {M.}~\bibnamefont
  {Cotrufo}}, \bibinfo {author} {\bibfnamefont {J.}~\bibnamefont {Kullig}},
  \bibinfo {author} {\bibfnamefont {J.}~\bibnamefont {Wiersig}},\ and\ \bibinfo
  {author} {\bibfnamefont {A.}~\bibnamefont {Al{\`u}}},\ }\bibfield  {title}
  {\bibinfo {title} {Coherent control of chaotic optical microcavity with
  reflectionless scattering modes},\ }\href
  {https://doi.org/10.1038/s41567-023-02242-w} {\bibfield  {journal} {\bibinfo
  {journal} {Nat. Phys.}\ }\textbf {\bibinfo {volume} {20}},\ \bibinfo {pages}
  {109} (\bibinfo {year} {2024})}\BibitemShut {NoStop}%
\bibitem [{\citenamefont {Xiao}\ \emph {et~al.}(2013)\citenamefont {Xiao},
  \citenamefont {Jiang}, \citenamefont {Yang}, \citenamefont {Wang},
  \citenamefont {Shi}, \citenamefont {Li},\ and\ \citenamefont
  {Gong}}]{xiaoTunnelinginducedTransparencyChaotic2013}%
  \BibitemOpen
  \bibfield  {author} {\bibinfo {author} {\bibfnamefont {Y.-F.}\ \bibnamefont
  {Xiao}}, \bibinfo {author} {\bibfnamefont {X.-F.}\ \bibnamefont {Jiang}},
  \bibinfo {author} {\bibfnamefont {Q.-F.}\ \bibnamefont {Yang}}, \bibinfo
  {author} {\bibfnamefont {L.}~\bibnamefont {Wang}}, \bibinfo {author}
  {\bibfnamefont {K.}~\bibnamefont {Shi}}, \bibinfo {author} {\bibfnamefont
  {Y.}~\bibnamefont {Li}},\ and\ \bibinfo {author} {\bibfnamefont
  {Q.}~\bibnamefont {Gong}},\ }\bibfield  {title} {\bibinfo {title}
  {Tunneling-induced transparency in a chaotic microcavity},\ }\href
  {https://doi.org/10.1002/lpor.201300042} {\bibfield  {journal} {\bibinfo
  {journal} {Laser Photonics Rev.}\ }\textbf {\bibinfo {volume} {7}},\ \bibinfo
  {pages} {L51} (\bibinfo {year} {2013})}\BibitemShut {NoStop}%
\bibitem [{\citenamefont {Hayes}\ \emph {et~al.}(1993)\citenamefont {Hayes},
  \citenamefont {Grebogi},\ and\ \citenamefont
  {Ott}}]{hayesCommunicatingChaos1993}%
  \BibitemOpen
  \bibfield  {author} {\bibinfo {author} {\bibfnamefont {S.}~\bibnamefont
  {Hayes}}, \bibinfo {author} {\bibfnamefont {C.}~\bibnamefont {Grebogi}},\
  and\ \bibinfo {author} {\bibfnamefont {E.}~\bibnamefont {Ott}},\ }\bibfield
  {title} {\bibinfo {title} {Communicating with chaos},\ }\href
  {https://doi.org/10.1103/PhysRevLett.70.3031} {\bibfield  {journal} {\bibinfo
   {journal} {Phys. Rev. Lett.}\ }\textbf {\bibinfo {volume} {70}},\ \bibinfo
  {pages} {3031} (\bibinfo {year} {1993})}\BibitemShut {NoStop}%
\bibitem [{\citenamefont {VanWiggeren}\ and\ \citenamefont
  {Roy}(1998)}]{vanwiggerenCommunicationChaoticLasers1998}%
  \BibitemOpen
  \bibfield  {author} {\bibinfo {author} {\bibfnamefont {G.~D.}\ \bibnamefont
  {VanWiggeren}}\ and\ \bibinfo {author} {\bibfnamefont {R.}~\bibnamefont
  {Roy}},\ }\bibfield  {title} {\bibinfo {title} {Communication with chaotic
  lasers},\ }\href {https://doi.org/10.1126/science.279.5354.1198} {\bibfield
  {journal} {\bibinfo  {journal} {Science}\ }\textbf {\bibinfo {volume}
  {279}},\ \bibinfo {pages} {1198} (\bibinfo {year} {1998})}\BibitemShut
  {NoStop}%
\bibitem [{\citenamefont {Uchida}\ \emph {et~al.}(2008)\citenamefont {Uchida},
  \citenamefont {Amano}, \citenamefont {Inoue}, \citenamefont {Hirano},
  \citenamefont {Naito}, \citenamefont {Someya}, \citenamefont {Oowada},
  \citenamefont {Kurashige}, \citenamefont {Shiki}, \citenamefont {Yoshimori},
  \citenamefont {Yoshimura},\ and\ \citenamefont
  {Davis}}]{uchidaFastPhysicalRandom2008}%
  \BibitemOpen
  \bibfield  {author} {\bibinfo {author} {\bibfnamefont {A.}~\bibnamefont
  {Uchida}}, \bibinfo {author} {\bibfnamefont {K.}~\bibnamefont {Amano}},
  \bibinfo {author} {\bibfnamefont {M.}~\bibnamefont {Inoue}}, \bibinfo
  {author} {\bibfnamefont {K.}~\bibnamefont {Hirano}}, \bibinfo {author}
  {\bibfnamefont {S.}~\bibnamefont {Naito}}, \bibinfo {author} {\bibfnamefont
  {H.}~\bibnamefont {Someya}}, \bibinfo {author} {\bibfnamefont
  {I.}~\bibnamefont {Oowada}}, \bibinfo {author} {\bibfnamefont
  {T.}~\bibnamefont {Kurashige}}, \bibinfo {author} {\bibfnamefont
  {M.}~\bibnamefont {Shiki}}, \bibinfo {author} {\bibfnamefont
  {S.}~\bibnamefont {Yoshimori}}, \bibinfo {author} {\bibfnamefont
  {K.}~\bibnamefont {Yoshimura}},\ and\ \bibinfo {author} {\bibfnamefont
  {P.}~\bibnamefont {Davis}},\ }\bibfield  {title} {\bibinfo {title} {Fast
  physical random bit generation with chaotic semiconductor lasers},\ }\href
  {https://doi.org/10.1038/nphoton.2008.227} {\bibfield  {journal} {\bibinfo
  {journal} {Nat. Photonics}\ }\textbf {\bibinfo {volume} {2}},\ \bibinfo
  {pages} {728} (\bibinfo {year} {2008})}\BibitemShut {NoStop}%
\bibitem [{\citenamefont {{Guanyu Wang}}\ \emph {et~al.}(1999)\citenamefont
  {{Guanyu Wang}}, \citenamefont {{Dajun Chen}}, \citenamefont {{Jianya Lin}},\
  and\ \citenamefont {{Xing
  Chen}}}]{guanyuwangApplicationChaoticOscillators1999}%
  \BibitemOpen
  \bibfield  {author} {\bibinfo {author} {\bibnamefont {{Guanyu Wang}}},
  \bibinfo {author} {\bibnamefont {{Dajun Chen}}}, \bibinfo {author}
  {\bibnamefont {{Jianya Lin}}},\ and\ \bibinfo {author} {\bibnamefont {{Xing
  Chen}}},\ }\bibfield  {title} {\bibinfo {title} {The application of chaotic
  oscillators to weak signal detection},\ }\href
  {https://doi.org/10.1109/41.753783} {\bibfield  {journal} {\bibinfo
  {journal} {IEEE Trans. Ind. Electron.}\ }\textbf {\bibinfo {volume} {46}},\
  \bibinfo {pages} {440} (\bibinfo {year} {1999})}\BibitemShut {NoStop}%
\bibitem [{\citenamefont {Carmon}\ \emph {et~al.}(2007)\citenamefont {Carmon},
  \citenamefont {Cross},\ and\ \citenamefont
  {Vahala}}]{carmonChaoticQuiveringMicronScaled2007}%
  \BibitemOpen
  \bibfield  {author} {\bibinfo {author} {\bibfnamefont {T.}~\bibnamefont
  {Carmon}}, \bibinfo {author} {\bibfnamefont {M.~C.}\ \bibnamefont {Cross}},\
  and\ \bibinfo {author} {\bibfnamefont {K.~J.}\ \bibnamefont {Vahala}},\
  }\bibfield  {title} {\bibinfo {title} {Chaotic quivering of micron-scaled
  on-chip resonators excited by centrifugal optical pressure},\ }\href
  {https://doi.org/10.1103/PhysRevLett.98.167203} {\bibfield  {journal}
  {\bibinfo  {journal} {Phys. Rev. Lett.}\ }\textbf {\bibinfo {volume} {98}},\
  \bibinfo {pages} {167203} (\bibinfo {year} {2007})}\BibitemShut {NoStop}%
\bibitem [{\citenamefont {Bakemeier}\ \emph {et~al.}(2015)\citenamefont
  {Bakemeier}, \citenamefont {Alvermann},\ and\ \citenamefont
  {Fehske}}]{bakemeierRouteChaosOptomechanics2015}%
  \BibitemOpen
  \bibfield  {author} {\bibinfo {author} {\bibfnamefont {L.}~\bibnamefont
  {Bakemeier}}, \bibinfo {author} {\bibfnamefont {A.}~\bibnamefont
  {Alvermann}},\ and\ \bibinfo {author} {\bibfnamefont {H.}~\bibnamefont
  {Fehske}},\ }\bibfield  {title} {\bibinfo {title} {Route to chaos in
  optomechanics},\ }\href {https://doi.org/10.1103/PhysRevLett.114.013601}
  {\bibfield  {journal} {\bibinfo  {journal} {Phys. Rev. Lett.}\ }\textbf
  {\bibinfo {volume} {114}},\ \bibinfo {pages} {013601} (\bibinfo {year}
  {2015})}\BibitemShut {NoStop}%
\bibitem [{\citenamefont {Monifi}\ \emph {et~al.}(2016)\citenamefont {Monifi},
  \citenamefont {Zhang}, \citenamefont {{\"O}zdemir}, \citenamefont {Peng},
  \citenamefont {Liu}, \citenamefont {Bo}, \citenamefont {Nori},\ and\
  \citenamefont {Yang}}]{monifiOptomechanicallyInducedStochastic2016a}%
  \BibitemOpen
  \bibfield  {author} {\bibinfo {author} {\bibfnamefont {F.}~\bibnamefont
  {Monifi}}, \bibinfo {author} {\bibfnamefont {J.}~\bibnamefont {Zhang}},
  \bibinfo {author} {\bibfnamefont {{\c S}.~K.}\ \bibnamefont {{\"O}zdemir}},
  \bibinfo {author} {\bibfnamefont {B.}~\bibnamefont {Peng}}, \bibinfo {author}
  {\bibfnamefont {Y.-x.}\ \bibnamefont {Liu}}, \bibinfo {author} {\bibfnamefont
  {F.}~\bibnamefont {Bo}}, \bibinfo {author} {\bibfnamefont {F.}~\bibnamefont
  {Nori}},\ and\ \bibinfo {author} {\bibfnamefont {L.}~\bibnamefont {Yang}},\
  }\bibfield  {title} {\bibinfo {title} {Optomechanically induced stochastic
  resonance and chaos transfer between optical fields},\ }\href
  {https://doi.org/10.1038/nphoton.2016.73} {\bibfield  {journal} {\bibinfo
  {journal} {Nat. Photonics}\ }\textbf {\bibinfo {volume} {10}},\ \bibinfo
  {pages} {399} (\bibinfo {year} {2016})}\BibitemShut {NoStop}%
\bibitem [{\citenamefont {Manneville}\ and\ \citenamefont
  {Pomeau}(1979)}]{mannevilleIntermittencyLorenzModel1979}%
  \BibitemOpen
  \bibfield  {author} {\bibinfo {author} {\bibfnamefont {P.}~\bibnamefont
  {Manneville}}\ and\ \bibinfo {author} {\bibfnamefont {Y.}~\bibnamefont
  {Pomeau}},\ }\bibfield  {title} {\bibinfo {title} {Intermittency and the
  {{Lorenz}} model},\ }\href {https://doi.org/10.1016/0375-9601(79)90255-X}
  {\bibfield  {journal} {\bibinfo  {journal} {Phys. Lett. A}\ }\textbf
  {\bibinfo {volume} {75}},\ \bibinfo {pages} {1} (\bibinfo {year}
  {1979})}\BibitemShut {NoStop}%
\bibitem [{\citenamefont {Pomeau}\ and\ \citenamefont
  {Manneville}(1980)}]{pomeauIntermittentTransitionTurbulencea}%
  \BibitemOpen
  \bibfield  {author} {\bibinfo {author} {\bibfnamefont {Y.}~\bibnamefont
  {Pomeau}}\ and\ \bibinfo {author} {\bibfnamefont {P.}~\bibnamefont
  {Manneville}},\ }\bibfield  {title} {\bibinfo {title} {Intermittent
  transition to turbulence in dissipative dynamical systems},\ }\href
  {https://doi.org/10.1007/BF01197757} {\bibfield  {journal} {\bibinfo
  {journal} {Commun. Math. Phys.}\ }\textbf {\bibinfo {volume} {74}},\ \bibinfo
  {pages} {189} (\bibinfo {year} {1980})}\BibitemShut {NoStop}%
\bibitem [{\citenamefont {Qin}\ \emph {et~al.}(1989)\citenamefont {Qin},
  \citenamefont {Wang}, \citenamefont {Yuan}, \citenamefont {Gao},\ and\
  \citenamefont {Zhang}}]{qinChaosBifurcationsPeriodic1989}%
  \BibitemOpen
  \bibfield  {author} {\bibinfo {author} {\bibfnamefont {J.}~\bibnamefont
  {Qin}}, \bibinfo {author} {\bibfnamefont {L.}~\bibnamefont {Wang}}, \bibinfo
  {author} {\bibfnamefont {D.~P.}\ \bibnamefont {Yuan}}, \bibinfo {author}
  {\bibfnamefont {P.}~\bibnamefont {Gao}},\ and\ \bibinfo {author}
  {\bibfnamefont {B.~Z.}\ \bibnamefont {Zhang}},\ }\bibfield  {title} {\bibinfo
  {title} {Chaos and bifurcations in periodic windows observed in plasmas},\
  }\href {https://doi.org/10.1103/PhysRevLett.63.163} {\bibfield  {journal}
  {\bibinfo  {journal} {Phys. Rev. Lett.}\ }\textbf {\bibinfo {volume} {63}},\
  \bibinfo {pages} {163} (\bibinfo {year} {1989})}\BibitemShut {NoStop}%
\bibitem [{\citenamefont {Chen}\ \emph {et~al.}(1994)\citenamefont {Chen},
  \citenamefont {Jiang},\ and\ \citenamefont
  {Fu}}]{fu-shengIntermittentChaosPulsating}%
  \BibitemOpen
  \bibfield  {author} {\bibinfo {author} {\bibfnamefont {F.-s.}\ \bibnamefont
  {Chen}}, \bibinfo {author} {\bibfnamefont {S.-y.}\ \bibnamefont {Jiang}},\
  and\ \bibinfo {author} {\bibfnamefont {J.-n.}\ \bibnamefont {Fu}},\
  }\bibfield  {title} {\bibinfo {title} {Intermittent chaos in pulsating
  stars},\ }\href {https://doi.org/10.1016/0275-1062(94)90042-6} {\bibfield
  {journal} {\bibinfo  {journal} {Chin. Astron. Astrophys.}\ }\textbf {\bibinfo
  {volume} {18}},\ \bibinfo {pages} {292} (\bibinfo {year} {1994})}\BibitemShut
  {NoStop}%
\bibitem [{\citenamefont {Chernin}\ and\ \citenamefont
  {Valtonen}(1998)}]{cherninIntermittentChaosThreebody1998}%
  \BibitemOpen
  \bibfield  {author} {\bibinfo {author} {\bibfnamefont {A.~D.}\ \bibnamefont
  {Chernin}}\ and\ \bibinfo {author} {\bibfnamefont {M.~J.}\ \bibnamefont
  {Valtonen}},\ }\bibfield  {title} {\bibinfo {title} {Intermittent chaos in
  three-body dynamics},\ }\href {https://doi.org/10.1016/S1387-6473(98)00021-9}
  {\bibfield  {journal} {\bibinfo  {journal} {New Astron. Rev.}\ }\textbf
  {\bibinfo {volume} {42}},\ \bibinfo {pages} {41} (\bibinfo {year}
  {1998})}\BibitemShut {NoStop}%
\bibitem [{\citenamefont {Hramov}\ \emph {et~al.}(2015)\citenamefont {Hramov},
  \citenamefont {Makarov}, \citenamefont {Maximenko}, \citenamefont
  {Koronovskii},\ and\ \citenamefont
  {Balanov}}]{hramovIntermittencyRouteChaos2015}%
  \BibitemOpen
  \bibfield  {author} {\bibinfo {author} {\bibfnamefont {A.~E.}\ \bibnamefont
  {Hramov}}, \bibinfo {author} {\bibfnamefont {V.~V.}\ \bibnamefont {Makarov}},
  \bibinfo {author} {\bibfnamefont {V.~A.}\ \bibnamefont {Maximenko}}, \bibinfo
  {author} {\bibfnamefont {A.~A.}\ \bibnamefont {Koronovskii}},\ and\ \bibinfo
  {author} {\bibfnamefont {A.~G.}\ \bibnamefont {Balanov}},\ }\bibfield
  {title} {\bibinfo {title} {Intermittency route to chaos and broadband
  high-frequency generation in semiconductor superlattice coupled to external
  resonator},\ }\href {https://doi.org/10.1103/PhysRevE.92.022911} {\bibfield
  {journal} {\bibinfo  {journal} {Phys. Rev. E}\ }\textbf {\bibinfo {volume}
  {92}},\ \bibinfo {pages} {022911} (\bibinfo {year} {2015})}\BibitemShut
  {NoStop}%
\bibitem [{\citenamefont {Suzuki}\ \emph {et~al.}(2016)\citenamefont {Suzuki},
  \citenamefont {Lu}, \citenamefont {Ben-Jacob},\ and\ \citenamefont
  {Onuchic}}]{suzukiPeriodicQuasiperiodicChaotic2016}%
  \BibitemOpen
  \bibfield  {author} {\bibinfo {author} {\bibfnamefont {Y.}~\bibnamefont
  {Suzuki}}, \bibinfo {author} {\bibfnamefont {M.}~\bibnamefont {Lu}}, \bibinfo
  {author} {\bibfnamefont {E.}~\bibnamefont {Ben-Jacob}},\ and\ \bibinfo
  {author} {\bibfnamefont {J.~N.}\ \bibnamefont {Onuchic}},\ }\bibfield
  {title} {\bibinfo {title} {Periodic, quasi-periodic and chaotic dynamics in
  simple gene elements with time delays},\ }\href
  {https://doi.org/10.1038/srep21037} {\bibfield  {journal} {\bibinfo
  {journal} {Sci. Rep.}\ }\textbf {\bibinfo {volume} {6}},\ \bibinfo {pages}
  {21037} (\bibinfo {year} {2016})}\BibitemShut {NoStop}%
\bibitem [{\citenamefont {Bubanja}\ \emph {et~al.}(2016)\citenamefont
  {Bubanja}, \citenamefont {Ma{\'c}e{\v s}i{\'c}}, \citenamefont
  {{Ivanovi{\'c}-{\v S}a{\v s}i{\'c}}}, \citenamefont {{\v C}upi{\'c}},
  \citenamefont {Ani{\'c}},\ and\ \citenamefont
  {{Kolar-Ani{\'c}}}}]{bubanjaIntermittentChaosBray2016}%
  \BibitemOpen
  \bibfield  {author} {\bibinfo {author} {\bibfnamefont {I.~N.}\ \bibnamefont
  {Bubanja}}, \bibinfo {author} {\bibfnamefont {S.}~\bibnamefont {Ma{\'c}e{\v
  s}i{\'c}}}, \bibinfo {author} {\bibfnamefont {A.}~\bibnamefont
  {{Ivanovi{\'c}-{\v S}a{\v s}i{\'c}}}}, \bibinfo {author} {\bibfnamefont {{\v
  Z}.}~\bibnamefont {{\v C}upi{\'c}}}, \bibinfo {author} {\bibfnamefont
  {S.}~\bibnamefont {Ani{\'c}}},\ and\ \bibinfo {author} {\bibfnamefont
  {{\relax Lj}.}~\bibnamefont {{Kolar-Ani{\'c}}}},\ }\bibfield  {title}
  {\bibinfo {title} {Intermittent chaos in the {{Bray}}--{{Liebhafsky}}
  oscillator. {{Temperature}} dependence},\ }\href
  {https://doi.org/10.1039/C6CP00759G} {\bibfield  {journal} {\bibinfo
  {journal} {Phys. Chem. Chem. Phys.}\ }\textbf {\bibinfo {volume} {18}},\
  \bibinfo {pages} {9770} (\bibinfo {year} {2016})}\BibitemShut {NoStop}%
\bibitem [{\citenamefont {Aizawa}(1983)}]{aizawaSymbolicDynamicsApproach1983}%
  \BibitemOpen
  \bibfield  {author} {\bibinfo {author} {\bibfnamefont {Y.}~\bibnamefont
  {Aizawa}},\ }\bibfield  {title} {\bibinfo {title} {Symbolic dynamics approach
  to intermittent chaos},\ }\href {https://doi.org/10.1143/PTP.70.1249}
  {\bibfield  {journal} {\bibinfo  {journal} {Prog. Theor. Phys.}\ }\textbf
  {\bibinfo {volume} {70}},\ \bibinfo {pages} {1249} (\bibinfo {year}
  {1983})}\BibitemShut {NoStop}%
\bibitem [{\citenamefont {Brunton}\ \emph {et~al.}(2017)\citenamefont
  {Brunton}, \citenamefont {Brunton}, \citenamefont {Proctor}, \citenamefont
  {Kaiser},\ and\ \citenamefont {Kutz}}]{bruntonChaosIntermittentlyForced2017}%
  \BibitemOpen
  \bibfield  {author} {\bibinfo {author} {\bibfnamefont {S.~L.}\ \bibnamefont
  {Brunton}}, \bibinfo {author} {\bibfnamefont {B.~W.}\ \bibnamefont
  {Brunton}}, \bibinfo {author} {\bibfnamefont {J.~L.}\ \bibnamefont
  {Proctor}}, \bibinfo {author} {\bibfnamefont {E.}~\bibnamefont {Kaiser}},\
  and\ \bibinfo {author} {\bibfnamefont {J.~N.}\ \bibnamefont {Kutz}},\
  }\bibfield  {title} {\bibinfo {title} {Chaos as an intermittently forced
  linear system},\ }\href {https://doi.org/10.1038/s41467-017-00030-8}
  {\bibfield  {journal} {\bibinfo  {journal} {Nat. Commun.}\ }\textbf {\bibinfo
  {volume} {8}},\ \bibinfo {pages} {19} (\bibinfo {year} {2017})}\BibitemShut
  {NoStop}%
\bibitem [{\citenamefont {Zhang}\ \emph {et~al.}(2020)\citenamefont {Zhang},
  \citenamefont {You},\ and\ \citenamefont
  {L{\"u}}}]{zhangIntermittentChaosCavity2020}%
  \BibitemOpen
  \bibfield  {author} {\bibinfo {author} {\bibfnamefont {D.-W.}\ \bibnamefont
  {Zhang}}, \bibinfo {author} {\bibfnamefont {C.}~\bibnamefont {You}},\ and\
  \bibinfo {author} {\bibfnamefont {X.-Y.}\ \bibnamefont {L{\"u}}},\ }\bibfield
   {title} {\bibinfo {title} {Intermittent chaos in cavity optomechanics},\
  }\href {https://doi.org/10.1103/PhysRevA.101.053851} {\bibfield  {journal}
  {\bibinfo  {journal} {Phys. Rev. A}\ }\textbf {\bibinfo {volume} {101}},\
  \bibinfo {pages} {053851} (\bibinfo {year} {2020})}\BibitemShut {NoStop}%
\bibitem [{\citenamefont {Benzi}\ \emph {et~al.}(1986)\citenamefont {Benzi},
  \citenamefont {Paladin}, \citenamefont {Patarnello}, \citenamefont
  {Santangelo},\ and\ \citenamefont
  {Vulpiani}}]{benziIntermittencyCoherentStructures1986}%
  \BibitemOpen
  \bibfield  {author} {\bibinfo {author} {\bibfnamefont {R.}~\bibnamefont
  {Benzi}}, \bibinfo {author} {\bibfnamefont {G.}~\bibnamefont {Paladin}},
  \bibinfo {author} {\bibfnamefont {S.}~\bibnamefont {Patarnello}}, \bibinfo
  {author} {\bibfnamefont {P.}~\bibnamefont {Santangelo}},\ and\ \bibinfo
  {author} {\bibfnamefont {A.}~\bibnamefont {Vulpiani}},\ }\bibfield  {title}
  {\bibinfo {title} {Intermittency and coherent structures in two-dimensional
  turbulence},\ }\href {https://doi.org/10.1088/0305-4470/19/18/023} {\bibfield
   {journal} {\bibinfo  {journal} {J. Phys. A: Math. Gen.}\ }\textbf {\bibinfo
  {volume} {19}},\ \bibinfo {pages} {3771} (\bibinfo {year}
  {1986})}\BibitemShut {NoStop}%
\bibitem [{\citenamefont {Tang}\ \emph {et~al.}(1992)\citenamefont {Tang},
  \citenamefont {Li},\ and\ \citenamefont
  {Weiss}}]{tangLaserDynamicsTypeI1992}%
  \BibitemOpen
  \bibfield  {author} {\bibinfo {author} {\bibfnamefont {D.~Y.}\ \bibnamefont
  {Tang}}, \bibinfo {author} {\bibfnamefont {M.~Y.}\ \bibnamefont {Li}},\ and\
  \bibinfo {author} {\bibfnamefont {C.~O.}\ \bibnamefont {Weiss}},\ }\bibfield
  {title} {\bibinfo {title} {Laser dynamics of type-{{I}} intermittency},\
  }\href {https://doi.org/10.1103/PhysRevA.46.676} {\bibfield  {journal}
  {\bibinfo  {journal} {Phys. Rev. A}\ }\textbf {\bibinfo {volume} {46}},\
  \bibinfo {pages} {676} (\bibinfo {year} {1992})}\BibitemShut {NoStop}%
\bibitem [{\citenamefont {Shirahama}\ \emph {et~al.}(1998)\citenamefont
  {Shirahama}, \citenamefont {Fukushima}, \citenamefont {Yoshida},\ and\
  \citenamefont {Taniguchi}}]{shirahamaIntermittentChaosMutually1998}%
  \BibitemOpen
  \bibfield  {author} {\bibinfo {author} {\bibfnamefont {H.}~\bibnamefont
  {Shirahama}}, \bibinfo {author} {\bibfnamefont {K.}~\bibnamefont
  {Fukushima}}, \bibinfo {author} {\bibfnamefont {N.}~\bibnamefont {Yoshida}},\
  and\ \bibinfo {author} {\bibfnamefont {K.}~\bibnamefont {Taniguchi}},\
  }\bibfield  {title} {\bibinfo {title} {Intermittent chaos in a mutually
  coupled {{PLL}}'s system},\ }\href {https://doi.org/10.1109/81.728868}
  {\bibfield  {journal} {\bibinfo  {journal} {IEEE Trans. Circuits Syst. I:
  Fundam. Theory Appl.}\ }\textbf {\bibinfo {volume} {45}},\ \bibinfo {pages}
  {1114} (\bibinfo {year} {1998})}\BibitemShut {NoStop}%
\bibitem [{\citenamefont
  {Hoppensteadt}(1989)}]{hoppensteadtIntermittentChaosSelforganization1989}%
  \BibitemOpen
  \bibfield  {author} {\bibinfo {author} {\bibfnamefont {F.~C.}\ \bibnamefont
  {Hoppensteadt}},\ }\bibfield  {title} {\bibinfo {title} {Intermittent chaos,
  self-organization, and learning from synchronous synaptic activity in model
  neuron networks.},\ }\href {https://doi.org/10.1073/pnas.86.9.2991}
  {\bibfield  {journal} {\bibinfo  {journal} {Proc. Natl. Acad. Sci. U.S.A.}\
  }\textbf {\bibinfo {volume} {86}},\ \bibinfo {pages} {2991} (\bibinfo {year}
  {1989})}\BibitemShut {NoStop}%
\bibitem [{\citenamefont {Wojtusiak}\ \emph {et~al.}(2021)\citenamefont
  {Wojtusiak}, \citenamefont {Balanov},\ and\ \citenamefont
  {Savel'ev}}]{wojtusiakIntermittentMetastableChaos2021}%
  \BibitemOpen
  \bibfield  {author} {\bibinfo {author} {\bibfnamefont {A.}~\bibnamefont
  {Wojtusiak}}, \bibinfo {author} {\bibfnamefont {A.}~\bibnamefont {Balanov}},\
  and\ \bibinfo {author} {\bibfnamefont {S.}~\bibnamefont {Savel'ev}},\
  }\bibfield  {title} {\bibinfo {title} {Intermittent and metastable chaos in a
  memristive artificial neuron with inertia},\ }\href
  {https://doi.org/10.1016/j.chaos.2020.110383} {\bibfield  {journal} {\bibinfo
   {journal} {Chaos, Solitons Fractals}\ }\textbf {\bibinfo {volume} {142}},\
  \bibinfo {pages} {110383} (\bibinfo {year} {2021})}\BibitemShut {NoStop}%
\bibitem [{\citenamefont {Bosco}\ \emph {et~al.}(2017)\citenamefont {Bosco},
  \citenamefont {Sato}, \citenamefont {Terashima}, \citenamefont {Ohara},
  \citenamefont {Uchida}, \citenamefont {Harayama},\ and\ \citenamefont
  {Inubushi}}]{boscoRandomNumberGeneration2017}%
  \BibitemOpen
  \bibfield  {author} {\bibinfo {author} {\bibfnamefont {A.~K.~D.}\
  \bibnamefont {Bosco}}, \bibinfo {author} {\bibfnamefont {N.}~\bibnamefont
  {Sato}}, \bibinfo {author} {\bibfnamefont {Y.}~\bibnamefont {Terashima}},
  \bibinfo {author} {\bibfnamefont {S.}~\bibnamefont {Ohara}}, \bibinfo
  {author} {\bibfnamefont {A.}~\bibnamefont {Uchida}}, \bibinfo {author}
  {\bibfnamefont {T.}~\bibnamefont {Harayama}},\ and\ \bibinfo {author}
  {\bibfnamefont {M.}~\bibnamefont {Inubushi}},\ }\bibfield  {title} {\bibinfo
  {title} {Random number generation from intermittent optical chaos},\ }\href
  {https://doi.org/10.1109/JSTQE.2017.2708608} {\bibfield  {journal} {\bibinfo
  {journal} {IEEE J. Sel. Top. Quantum Electron.}\ }\textbf {\bibinfo {volume}
  {23}},\ \bibinfo {pages} {1} (\bibinfo {year} {2017})}\BibitemShut {NoStop}%
\bibitem [{\citenamefont {Armani}\ \emph {et~al.}(2003)\citenamefont {Armani},
  \citenamefont {Kippenberg}, \citenamefont {Spillane},\ and\ \citenamefont
  {Vahala}}]{armaniUltrahighQToroidMicrocavity2003}%
  \BibitemOpen
  \bibfield  {author} {\bibinfo {author} {\bibfnamefont {D.~K.}\ \bibnamefont
  {Armani}}, \bibinfo {author} {\bibfnamefont {T.~J.}\ \bibnamefont
  {Kippenberg}}, \bibinfo {author} {\bibfnamefont {S.~M.}\ \bibnamefont
  {Spillane}},\ and\ \bibinfo {author} {\bibfnamefont {K.~J.}\ \bibnamefont
  {Vahala}},\ }\bibfield  {title} {\bibinfo {title} {Ultra-high-{{Q}} toroid
  microcavity on a chip},\ }\href {https://doi.org/10.1038/nature01371}
  {\bibfield  {journal} {\bibinfo  {journal} {Nature}\ }\textbf {\bibinfo
  {volume} {421}},\ \bibinfo {pages} {925} (\bibinfo {year}
  {2003})}\BibitemShut {NoStop}%
\bibitem [{\citenamefont {Chen}\ \emph {et~al.}(2017)\citenamefont {Chen},
  \citenamefont {Kaya~{\"O}zdemir}, \citenamefont {Zhao}, \citenamefont
  {Wiersig},\ and\ \citenamefont {Yang}}]{chenExceptionalPointsEnhance2017}%
  \BibitemOpen
  \bibfield  {author} {\bibinfo {author} {\bibfnamefont {W.}~\bibnamefont
  {Chen}}, \bibinfo {author} {\bibfnamefont {{\c S}.}~\bibnamefont
  {Kaya~{\"O}zdemir}}, \bibinfo {author} {\bibfnamefont {G.}~\bibnamefont
  {Zhao}}, \bibinfo {author} {\bibfnamefont {J.}~\bibnamefont {Wiersig}},\ and\
  \bibinfo {author} {\bibfnamefont {L.}~\bibnamefont {Yang}},\ }\bibfield
  {title} {\bibinfo {title} {Exceptional points enhance sensing in an optical
  microcavity},\ }\href {https://doi.org/10.1038/nature23281} {\bibfield
  {journal} {\bibinfo  {journal} {Nature}\ }\textbf {\bibinfo {volume} {548}},\
  \bibinfo {pages} {192} (\bibinfo {year} {2017})}\BibitemShut {NoStop}%
\bibitem [{\citenamefont {Zhu}\ \emph {et~al.}(2010)\citenamefont {Zhu},
  \citenamefont {Ozdemir}, \citenamefont {Xiao}, \citenamefont {Li},
  \citenamefont {He}, \citenamefont {Chen},\ and\ \citenamefont
  {Yang}}]{zhuOnchipSingleNanoparticle2010a}%
  \BibitemOpen
  \bibfield  {author} {\bibinfo {author} {\bibfnamefont {J.}~\bibnamefont
  {Zhu}}, \bibinfo {author} {\bibfnamefont {S.~K.}\ \bibnamefont {Ozdemir}},
  \bibinfo {author} {\bibfnamefont {Y.-F.}\ \bibnamefont {Xiao}}, \bibinfo
  {author} {\bibfnamefont {L.}~\bibnamefont {Li}}, \bibinfo {author}
  {\bibfnamefont {L.}~\bibnamefont {He}}, \bibinfo {author} {\bibfnamefont
  {D.-R.}\ \bibnamefont {Chen}},\ and\ \bibinfo {author} {\bibfnamefont
  {L.}~\bibnamefont {Yang}},\ }\bibfield  {title} {\bibinfo {title} {On-chip
  single nanoparticle detection and sizing by mode splitting in an
  ultrahigh-{{Q}} microresonator},\ }\href
  {https://doi.org/10.1038/nphoton.2009.237} {\bibfield  {journal} {\bibinfo
  {journal} {Nat. Photonics}\ }\textbf {\bibinfo {volume} {4}},\ \bibinfo
  {pages} {46} (\bibinfo {year} {2010})}\BibitemShut {NoStop}%
\bibitem [{\citenamefont {Goldwin}\ \emph {et~al.}(2011)\citenamefont
  {Goldwin}, \citenamefont {Trupke}, \citenamefont {Kenner}, \citenamefont
  {Ratnapala},\ and\ \citenamefont
  {Hinds}}]{goldwinFastCavityenhancedAtom2011}%
  \BibitemOpen
  \bibfield  {author} {\bibinfo {author} {\bibfnamefont {J.}~\bibnamefont
  {Goldwin}}, \bibinfo {author} {\bibfnamefont {M.}~\bibnamefont {Trupke}},
  \bibinfo {author} {\bibfnamefont {J.}~\bibnamefont {Kenner}}, \bibinfo
  {author} {\bibfnamefont {A.}~\bibnamefont {Ratnapala}},\ and\ \bibinfo
  {author} {\bibfnamefont {E.}~\bibnamefont {Hinds}},\ }\bibfield  {title}
  {\bibinfo {title} {Fast cavity-enhanced atom detection with low noise and
  high fidelity},\ }\href {https://doi.org/10.1038/ncomms1428} {\bibfield
  {journal} {\bibinfo  {journal} {Nat. Commun.}\ }\textbf {\bibinfo {volume}
  {2}},\ \bibinfo {pages} {418} (\bibinfo {year} {2011})}\BibitemShut {NoStop}%
\bibitem [{\citenamefont {Su}\ \emph {et~al.}(2016)\citenamefont {Su},
  \citenamefont {Goldberg},\ and\ \citenamefont
  {Stoltz}}]{suLabelfreeDetectionSingle2016}%
  \BibitemOpen
  \bibfield  {author} {\bibinfo {author} {\bibfnamefont {J.}~\bibnamefont
  {Su}}, \bibinfo {author} {\bibfnamefont {A.~F.}\ \bibnamefont {Goldberg}},\
  and\ \bibinfo {author} {\bibfnamefont {B.~M.}\ \bibnamefont {Stoltz}},\
  }\bibfield  {title} {\bibinfo {title} {Label-free detection of single
  nanoparticles and biological molecules using microtoroid optical
  resonators},\ }\href {https://doi.org/10.1038/lsa.2016.1} {\bibfield
  {journal} {\bibinfo  {journal} {Light Sci. Appl.}\ }\textbf {\bibinfo
  {volume} {5}},\ \bibinfo {pages} {e16001} (\bibinfo {year}
  {2016})}\BibitemShut {NoStop}%
\bibitem [{\citenamefont {Li}\ \emph {et~al.}(2018)\citenamefont {Li},
  \citenamefont {B{\'i}lek}, \citenamefont {Hoff}, \citenamefont {Madsen},
  \citenamefont {Forstner}, \citenamefont {Prakash}, \citenamefont
  {Sch{\"a}fermeier}, \citenamefont {Gehring}, \citenamefont {Bowen},\ and\
  \citenamefont {Andersen}}]{liQuantumEnhancedOptomechanical2018}%
  \BibitemOpen
  \bibfield  {author} {\bibinfo {author} {\bibfnamefont {B.-B.}\ \bibnamefont
  {Li}}, \bibinfo {author} {\bibfnamefont {J.}~\bibnamefont {B{\'i}lek}},
  \bibinfo {author} {\bibfnamefont {U.~B.}\ \bibnamefont {Hoff}}, \bibinfo
  {author} {\bibfnamefont {L.~S.}\ \bibnamefont {Madsen}}, \bibinfo {author}
  {\bibfnamefont {S.}~\bibnamefont {Forstner}}, \bibinfo {author}
  {\bibfnamefont {V.}~\bibnamefont {Prakash}}, \bibinfo {author} {\bibfnamefont
  {C.}~\bibnamefont {Sch{\"a}fermeier}}, \bibinfo {author} {\bibfnamefont
  {T.}~\bibnamefont {Gehring}}, \bibinfo {author} {\bibfnamefont {W.~P.}\
  \bibnamefont {Bowen}},\ and\ \bibinfo {author} {\bibfnamefont {U.~L.}\
  \bibnamefont {Andersen}},\ }\bibfield  {title} {\bibinfo {title} {Quantum
  enhanced optomechanical magnetometry},\ }\href
  {https://doi.org/10.1364/OPTICA.5.000850} {\bibfield  {journal} {\bibinfo
  {journal} {Optica}\ }\textbf {\bibinfo {volume} {5}},\ \bibinfo {pages} {850}
  (\bibinfo {year} {2018})}\BibitemShut {NoStop}%
\bibitem [{\citenamefont {Forstner}\ \emph {et~al.}(2012)\citenamefont
  {Forstner}, \citenamefont {Prams}, \citenamefont {Knittel}, \citenamefont
  {{van Ooijen}}, \citenamefont {Swaim}, \citenamefont {Harris}, \citenamefont
  {Szorkovszky}, \citenamefont {Bowen},\ and\ \citenamefont
  {{Rubinsztein-Dunlop}}}]{forstnerCavityOptomechanicalMagnetometer2012}%
  \BibitemOpen
  \bibfield  {author} {\bibinfo {author} {\bibfnamefont {S.}~\bibnamefont
  {Forstner}}, \bibinfo {author} {\bibfnamefont {S.}~\bibnamefont {Prams}},
  \bibinfo {author} {\bibfnamefont {J.}~\bibnamefont {Knittel}}, \bibinfo
  {author} {\bibfnamefont {E.~D.}\ \bibnamefont {{van Ooijen}}}, \bibinfo
  {author} {\bibfnamefont {J.~D.}\ \bibnamefont {Swaim}}, \bibinfo {author}
  {\bibfnamefont {G.~I.}\ \bibnamefont {Harris}}, \bibinfo {author}
  {\bibfnamefont {A.}~\bibnamefont {Szorkovszky}}, \bibinfo {author}
  {\bibfnamefont {W.~P.}\ \bibnamefont {Bowen}},\ and\ \bibinfo {author}
  {\bibfnamefont {H.}~\bibnamefont {{Rubinsztein-Dunlop}}},\ }\bibfield
  {title} {\bibinfo {title} {Cavity optomechanical magnetometer},\ }\href
  {https://doi.org/10.1103/PhysRevLett.108.120801} {\bibfield  {journal}
  {\bibinfo  {journal} {Phys. Rev. Lett.}\ }\textbf {\bibinfo {volume} {108}},\
  \bibinfo {pages} {120801} (\bibinfo {year} {2012})}\BibitemShut {NoStop}%
\bibitem [{\citenamefont {{Basiri-Esfahani}}\ \emph {et~al.}(2019)\citenamefont
  {{Basiri-Esfahani}}, \citenamefont {Armin}, \citenamefont {Forstner},\ and\
  \citenamefont {Bowen}}]{basiri-esfahaniPrecisionUltrasoundSensing2019}%
  \BibitemOpen
  \bibfield  {author} {\bibinfo {author} {\bibfnamefont {S.}~\bibnamefont
  {{Basiri-Esfahani}}}, \bibinfo {author} {\bibfnamefont {A.}~\bibnamefont
  {Armin}}, \bibinfo {author} {\bibfnamefont {S.}~\bibnamefont {Forstner}},\
  and\ \bibinfo {author} {\bibfnamefont {W.~P.}\ \bibnamefont {Bowen}},\
  }\bibfield  {title} {\bibinfo {title} {Precision ultrasound sensing on a
  chip},\ }\href {https://doi.org/10.1038/s41467-018-08038-4} {\bibfield
  {journal} {\bibinfo  {journal} {Nat. Commun.}\ }\textbf {\bibinfo {volume}
  {10}},\ \bibinfo {pages} {132} (\bibinfo {year} {2019})}\BibitemShut
  {NoStop}%
\bibitem [{\citenamefont {Xu}\ \emph {et~al.}(2018)\citenamefont {Xu},
  \citenamefont {Chen}, \citenamefont {Zhao}, \citenamefont {Li}, \citenamefont
  {Lu},\ and\ \citenamefont
  {Yang}}]{xuWirelessWhisperinggallerymodeSensor2018}%
  \BibitemOpen
  \bibfield  {author} {\bibinfo {author} {\bibfnamefont {X.}~\bibnamefont
  {Xu}}, \bibinfo {author} {\bibfnamefont {W.}~\bibnamefont {Chen}}, \bibinfo
  {author} {\bibfnamefont {G.}~\bibnamefont {Zhao}}, \bibinfo {author}
  {\bibfnamefont {Y.}~\bibnamefont {Li}}, \bibinfo {author} {\bibfnamefont
  {C.}~\bibnamefont {Lu}},\ and\ \bibinfo {author} {\bibfnamefont
  {L.}~\bibnamefont {Yang}},\ }\bibfield  {title} {\bibinfo {title} {Wireless
  whispering-gallery-mode sensor for thermal sensing and aerial mapping},\
  }\href {https://doi.org/10.1038/s41377-018-0063-4} {\bibfield  {journal}
  {\bibinfo  {journal} {Light Sci. Appl.}\ }\textbf {\bibinfo {volume} {7}},\
  \bibinfo {pages} {62} (\bibinfo {year} {2018})}\BibitemShut {NoStop}%
\bibitem [{\citenamefont {Xu}\ \emph {et~al.}(2016)\citenamefont {Xu},
  \citenamefont {Jiang}, \citenamefont {Zhao},\ and\ \citenamefont
  {Yang}}]{xuPhonesizedWhisperinggalleryMicroresonator2016}%
  \BibitemOpen
  \bibfield  {author} {\bibinfo {author} {\bibfnamefont {X.}~\bibnamefont
  {Xu}}, \bibinfo {author} {\bibfnamefont {X.}~\bibnamefont {Jiang}}, \bibinfo
  {author} {\bibfnamefont {G.}~\bibnamefont {Zhao}},\ and\ \bibinfo {author}
  {\bibfnamefont {L.}~\bibnamefont {Yang}},\ }\bibfield  {title} {\bibinfo
  {title} {Phone-sized whispering-gallery microresonator sensing system},\
  }\href {https://doi.org/10.1364/OE.24.025905} {\bibfield  {journal} {\bibinfo
   {journal} {Opt. Express}\ }\textbf {\bibinfo {volume} {24}},\ \bibinfo
  {pages} {25905} (\bibinfo {year} {2016})}\BibitemShut {NoStop}%
\bibitem [{\citenamefont {Liao}\ and\ \citenamefont
  {Yang}(2021)}]{liaoOpticalWhisperinggalleryMode2021}%
  \BibitemOpen
  \bibfield  {author} {\bibinfo {author} {\bibfnamefont {J.}~\bibnamefont
  {Liao}}\ and\ \bibinfo {author} {\bibfnamefont {L.}~\bibnamefont {Yang}},\
  }\bibfield  {title} {\bibinfo {title} {Optical whispering-gallery mode
  barcodes for high-precision and wide-range temperature measurements},\ }\href
  {https://doi.org/10.1038/s41377-021-00472-2} {\bibfield  {journal} {\bibinfo
  {journal} {Light Sci. Appl.}\ }\textbf {\bibinfo {volume} {10}},\ \bibinfo
  {pages} {32} (\bibinfo {year} {2021})}\BibitemShut {NoStop}%
\bibitem [{\citenamefont {Rosenblum}\ \emph {et~al.}(2015)\citenamefont
  {Rosenblum}, \citenamefont {Lovsky}, \citenamefont {Arazi}, \citenamefont
  {Vollmer},\ and\ \citenamefont
  {Dayan}}]{rosenblumCavityRingupSpectroscopy2015}%
  \BibitemOpen
  \bibfield  {author} {\bibinfo {author} {\bibfnamefont {S.}~\bibnamefont
  {Rosenblum}}, \bibinfo {author} {\bibfnamefont {Y.}~\bibnamefont {Lovsky}},
  \bibinfo {author} {\bibfnamefont {L.}~\bibnamefont {Arazi}}, \bibinfo
  {author} {\bibfnamefont {F.}~\bibnamefont {Vollmer}},\ and\ \bibinfo {author}
  {\bibfnamefont {B.}~\bibnamefont {Dayan}},\ }\bibfield  {title} {\bibinfo
  {title} {Cavity ring-up spectroscopy for ultrafast sensing with optical
  microresonators},\ }\href {https://doi.org/10.1038/ncomms7788} {\bibfield
  {journal} {\bibinfo  {journal} {Nat. Commun.}\ }\textbf {\bibinfo {volume}
  {6}},\ \bibinfo {pages} {6788} (\bibinfo {year} {2015})}\BibitemShut
  {NoStop}%
\bibitem [{\citenamefont {Lai}\ \emph {et~al.}(2019)\citenamefont {Lai},
  \citenamefont {Lu}, \citenamefont {Suh}, \citenamefont {Yuan},\ and\
  \citenamefont {Vahala}}]{laiObservationExceptionalpointenhancedSagnac2019a}%
  \BibitemOpen
  \bibfield  {author} {\bibinfo {author} {\bibfnamefont {Y.-H.}\ \bibnamefont
  {Lai}}, \bibinfo {author} {\bibfnamefont {Y.-K.}\ \bibnamefont {Lu}},
  \bibinfo {author} {\bibfnamefont {M.-G.}\ \bibnamefont {Suh}}, \bibinfo
  {author} {\bibfnamefont {Z.}~\bibnamefont {Yuan}},\ and\ \bibinfo {author}
  {\bibfnamefont {K.}~\bibnamefont {Vahala}},\ }\bibfield  {title} {\bibinfo
  {title} {Observation of the exceptional-point-enhanced {{Sagnac}} effect},\
  }\href {https://doi.org/10.1038/s41586-019-1777-z} {\bibfield  {journal}
  {\bibinfo  {journal} {Nature}\ }\textbf {\bibinfo {volume} {576}},\ \bibinfo
  {pages} {65} (\bibinfo {year} {2019})}\BibitemShut {NoStop}%
\bibitem [{\citenamefont {Lai}\ \emph {et~al.}(2020)\citenamefont {Lai},
  \citenamefont {Suh}, \citenamefont {Lu}, \citenamefont {Shen}, \citenamefont
  {Yang}, \citenamefont {Wang}, \citenamefont {Li}, \citenamefont {Lee},
  \citenamefont {Yang},\ and\ \citenamefont
  {Vahala}}]{laiEarthRotationMeasured2020a}%
  \BibitemOpen
  \bibfield  {author} {\bibinfo {author} {\bibfnamefont {Y.-H.}\ \bibnamefont
  {Lai}}, \bibinfo {author} {\bibfnamefont {M.-G.}\ \bibnamefont {Suh}},
  \bibinfo {author} {\bibfnamefont {Y.-K.}\ \bibnamefont {Lu}}, \bibinfo
  {author} {\bibfnamefont {B.}~\bibnamefont {Shen}}, \bibinfo {author}
  {\bibfnamefont {Q.-F.}\ \bibnamefont {Yang}}, \bibinfo {author}
  {\bibfnamefont {H.}~\bibnamefont {Wang}}, \bibinfo {author} {\bibfnamefont
  {J.}~\bibnamefont {Li}}, \bibinfo {author} {\bibfnamefont {S.~H.}\
  \bibnamefont {Lee}}, \bibinfo {author} {\bibfnamefont {K.~Y.}\ \bibnamefont
  {Yang}},\ and\ \bibinfo {author} {\bibfnamefont {K.}~\bibnamefont {Vahala}},\
  }\bibfield  {title} {\bibinfo {title} {Earth rotation measured by a
  chip-scale ring laser gyroscope},\ }\href
  {https://doi.org/10.1038/s41566-020-0588-y} {\bibfield  {journal} {\bibinfo
  {journal} {Nat. Photonics}\ }\textbf {\bibinfo {volume} {14}},\ \bibinfo
  {pages} {345} (\bibinfo {year} {2020})}\BibitemShut {NoStop}%
\bibitem [{\citenamefont {Qvarfort}\ \emph {et~al.}(2018)\citenamefont
  {Qvarfort}, \citenamefont {Serafini}, \citenamefont {Barker},\ and\
  \citenamefont {Bose}}]{qvarfortGravimetryNonlinearOptomechanics2018}%
  \BibitemOpen
  \bibfield  {author} {\bibinfo {author} {\bibfnamefont {S.}~\bibnamefont
  {Qvarfort}}, \bibinfo {author} {\bibfnamefont {A.}~\bibnamefont {Serafini}},
  \bibinfo {author} {\bibfnamefont {P.~F.}\ \bibnamefont {Barker}},\ and\
  \bibinfo {author} {\bibfnamefont {S.}~\bibnamefont {Bose}},\ }\bibfield
  {title} {\bibinfo {title} {Gravimetry through non-linear optomechanics},\
  }\href {https://doi.org/10.1038/s41467-018-06037-z} {\bibfield  {journal}
  {\bibinfo  {journal} {Nat. Commun.}\ }\textbf {\bibinfo {volume} {9}},\
  \bibinfo {pages} {3690} (\bibinfo {year} {2018})}\BibitemShut {NoStop}%
\bibitem [{\citenamefont {{Marin-Palomo}}\ \emph {et~al.}(2017)\citenamefont
  {{Marin-Palomo}}, \citenamefont {Kemal}, \citenamefont {Karpov},
  \citenamefont {Kordts}, \citenamefont {Pfeifle}, \citenamefont {Pfeiffer},
  \citenamefont {Trocha}, \citenamefont {Wolf}, \citenamefont {Brasch},
  \citenamefont {Anderson}, \citenamefont {Rosenberger}, \citenamefont
  {Vijayan}, \citenamefont {Freude}, \citenamefont {Kippenberg},\ and\
  \citenamefont {Koos}}]{marin-palomoMicroresonatorbasedSolitonsMassively2017}%
  \BibitemOpen
  \bibfield  {author} {\bibinfo {author} {\bibfnamefont {P.}~\bibnamefont
  {{Marin-Palomo}}}, \bibinfo {author} {\bibfnamefont {J.~N.}\ \bibnamefont
  {Kemal}}, \bibinfo {author} {\bibfnamefont {M.}~\bibnamefont {Karpov}},
  \bibinfo {author} {\bibfnamefont {A.}~\bibnamefont {Kordts}}, \bibinfo
  {author} {\bibfnamefont {J.}~\bibnamefont {Pfeifle}}, \bibinfo {author}
  {\bibfnamefont {M.~H.~P.}\ \bibnamefont {Pfeiffer}}, \bibinfo {author}
  {\bibfnamefont {P.}~\bibnamefont {Trocha}}, \bibinfo {author} {\bibfnamefont
  {S.}~\bibnamefont {Wolf}}, \bibinfo {author} {\bibfnamefont {V.}~\bibnamefont
  {Brasch}}, \bibinfo {author} {\bibfnamefont {M.~H.}\ \bibnamefont
  {Anderson}}, \bibinfo {author} {\bibfnamefont {R.}~\bibnamefont
  {Rosenberger}}, \bibinfo {author} {\bibfnamefont {K.}~\bibnamefont
  {Vijayan}}, \bibinfo {author} {\bibfnamefont {W.}~\bibnamefont {Freude}},
  \bibinfo {author} {\bibfnamefont {T.~J.}\ \bibnamefont {Kippenberg}},\ and\
  \bibinfo {author} {\bibfnamefont {C.}~\bibnamefont {Koos}},\ }\bibfield
  {title} {\bibinfo {title} {Microresonator-based solitons for massively
  parallel coherent optical communications},\ }\href
  {https://doi.org/10.1038/nature22387} {\bibfield  {journal} {\bibinfo
  {journal} {Nature}\ }\textbf {\bibinfo {volume} {546}},\ \bibinfo {pages}
  {274} (\bibinfo {year} {2017})}\BibitemShut {NoStop}%
\bibitem [{\citenamefont {Spencer}\ \emph {et~al.}(2018)\citenamefont
  {Spencer}, \citenamefont {Drake}, \citenamefont {Briles}, \citenamefont
  {Stone}, \citenamefont {Sinclair}, \citenamefont {Fredrick}, \citenamefont
  {Li}, \citenamefont {Westly}, \citenamefont {Ilic}, \citenamefont
  {Bluestone}, \citenamefont {Volet}, \citenamefont {Komljenovic},
  \citenamefont {Chang}, \citenamefont {Lee}, \citenamefont {Oh}, \citenamefont
  {Suh}, \citenamefont {Yang}, \citenamefont {Pfeiffer}, \citenamefont
  {Kippenberg}, \citenamefont {Norberg}, \citenamefont {Theogarajan},
  \citenamefont {Vahala}, \citenamefont {Newbury}, \citenamefont {Srinivasan},
  \citenamefont {Bowers}, \citenamefont {Diddams},\ and\ \citenamefont
  {Papp}}]{spencerOpticalfrequencySynthesizerUsing2018}%
  \BibitemOpen
  \bibfield  {author} {\bibinfo {author} {\bibfnamefont {D.~T.}\ \bibnamefont
  {Spencer}}, \bibinfo {author} {\bibfnamefont {T.}~\bibnamefont {Drake}},
  \bibinfo {author} {\bibfnamefont {T.~C.}\ \bibnamefont {Briles}}, \bibinfo
  {author} {\bibfnamefont {J.}~\bibnamefont {Stone}}, \bibinfo {author}
  {\bibfnamefont {L.~C.}\ \bibnamefont {Sinclair}}, \bibinfo {author}
  {\bibfnamefont {C.}~\bibnamefont {Fredrick}}, \bibinfo {author}
  {\bibfnamefont {Q.}~\bibnamefont {Li}}, \bibinfo {author} {\bibfnamefont
  {D.}~\bibnamefont {Westly}}, \bibinfo {author} {\bibfnamefont {B.~R.}\
  \bibnamefont {Ilic}}, \bibinfo {author} {\bibfnamefont {A.}~\bibnamefont
  {Bluestone}}, \bibinfo {author} {\bibfnamefont {N.}~\bibnamefont {Volet}},
  \bibinfo {author} {\bibfnamefont {T.}~\bibnamefont {Komljenovic}}, \bibinfo
  {author} {\bibfnamefont {L.}~\bibnamefont {Chang}}, \bibinfo {author}
  {\bibfnamefont {S.~H.}\ \bibnamefont {Lee}}, \bibinfo {author} {\bibfnamefont
  {D.~Y.}\ \bibnamefont {Oh}}, \bibinfo {author} {\bibfnamefont {M.-G.}\
  \bibnamefont {Suh}}, \bibinfo {author} {\bibfnamefont {K.~Y.}\ \bibnamefont
  {Yang}}, \bibinfo {author} {\bibfnamefont {M.~H.~P.}\ \bibnamefont
  {Pfeiffer}}, \bibinfo {author} {\bibfnamefont {T.~J.}\ \bibnamefont
  {Kippenberg}}, \bibinfo {author} {\bibfnamefont {E.}~\bibnamefont {Norberg}},
  \bibinfo {author} {\bibfnamefont {L.}~\bibnamefont {Theogarajan}}, \bibinfo
  {author} {\bibfnamefont {K.}~\bibnamefont {Vahala}}, \bibinfo {author}
  {\bibfnamefont {N.~R.}\ \bibnamefont {Newbury}}, \bibinfo {author}
  {\bibfnamefont {K.}~\bibnamefont {Srinivasan}}, \bibinfo {author}
  {\bibfnamefont {J.~E.}\ \bibnamefont {Bowers}}, \bibinfo {author}
  {\bibfnamefont {S.~A.}\ \bibnamefont {Diddams}},\ and\ \bibinfo {author}
  {\bibfnamefont {S.~B.}\ \bibnamefont {Papp}},\ }\bibfield  {title} {\bibinfo
  {title} {An optical-frequency synthesizer using integrated photonics},\
  }\href {https://doi.org/10.1038/s41586-018-0065-7} {\bibfield  {journal}
  {\bibinfo  {journal} {Nature}\ }\textbf {\bibinfo {volume} {557}},\ \bibinfo
  {pages} {81} (\bibinfo {year} {2018})}\BibitemShut {NoStop}%
\bibitem [{\citenamefont {Feldmann}\ \emph {et~al.}(2021)\citenamefont
  {Feldmann}, \citenamefont {Youngblood}, \citenamefont {Karpov}, \citenamefont
  {Gehring}, \citenamefont {Li}, \citenamefont {Stappers}, \citenamefont
  {Le~Gallo}, \citenamefont {Fu}, \citenamefont {Lukashchuk}, \citenamefont
  {Raja}, \citenamefont {Liu}, \citenamefont {Wright}, \citenamefont
  {Sebastian}, \citenamefont {Kippenberg}, \citenamefont {Pernice},\ and\
  \citenamefont {Bhaskaran}}]{feldmannParallelConvolutionalProcessing2021a}%
  \BibitemOpen
  \bibfield  {author} {\bibinfo {author} {\bibfnamefont {J.}~\bibnamefont
  {Feldmann}}, \bibinfo {author} {\bibfnamefont {N.}~\bibnamefont
  {Youngblood}}, \bibinfo {author} {\bibfnamefont {M.}~\bibnamefont {Karpov}},
  \bibinfo {author} {\bibfnamefont {H.}~\bibnamefont {Gehring}}, \bibinfo
  {author} {\bibfnamefont {X.}~\bibnamefont {Li}}, \bibinfo {author}
  {\bibfnamefont {M.}~\bibnamefont {Stappers}}, \bibinfo {author}
  {\bibfnamefont {M.}~\bibnamefont {Le~Gallo}}, \bibinfo {author}
  {\bibfnamefont {X.}~\bibnamefont {Fu}}, \bibinfo {author} {\bibfnamefont
  {A.}~\bibnamefont {Lukashchuk}}, \bibinfo {author} {\bibfnamefont {A.~S.}\
  \bibnamefont {Raja}}, \bibinfo {author} {\bibfnamefont {J.}~\bibnamefont
  {Liu}}, \bibinfo {author} {\bibfnamefont {C.~D.}\ \bibnamefont {Wright}},
  \bibinfo {author} {\bibfnamefont {A.}~\bibnamefont {Sebastian}}, \bibinfo
  {author} {\bibfnamefont {T.~J.}\ \bibnamefont {Kippenberg}}, \bibinfo
  {author} {\bibfnamefont {W.~H.~P.}\ \bibnamefont {Pernice}},\ and\ \bibinfo
  {author} {\bibfnamefont {H.}~\bibnamefont {Bhaskaran}},\ }\bibfield  {title}
  {\bibinfo {title} {Parallel convolutional processing using an integrated
  photonic tensor core},\ }\href {https://doi.org/10.1038/s41586-020-03070-1}
  {\bibfield  {journal} {\bibinfo  {journal} {Nature}\ }\textbf {\bibinfo
  {volume} {589}},\ \bibinfo {pages} {52} (\bibinfo {year} {2021})}\BibitemShut
  {NoStop}%
\bibitem [{\citenamefont {Aspelmeyer}\ \emph {et~al.}(2014)\citenamefont
  {Aspelmeyer}, \citenamefont {Kippenberg},\ and\ \citenamefont
  {Marquardt}}]{aspelmeyerCavityOptomechanics2014}%
  \BibitemOpen
  \bibfield  {author} {\bibinfo {author} {\bibfnamefont {M.}~\bibnamefont
  {Aspelmeyer}}, \bibinfo {author} {\bibfnamefont {T.~J.}\ \bibnamefont
  {Kippenberg}},\ and\ \bibinfo {author} {\bibfnamefont {F.}~\bibnamefont
  {Marquardt}},\ }\bibfield  {title} {\bibinfo {title} {Cavity optomechanics},\
  }\href {https://doi.org/10.1103/RevModPhys.86.1391} {\bibfield  {journal}
  {\bibinfo  {journal} {Rev. Mod. Phys.}\ }\textbf {\bibinfo {volume} {86}},\
  \bibinfo {pages} {1391} (\bibinfo {year} {2014})}\BibitemShut {NoStop}%
\bibitem [{\citenamefont {Grudinin}\ \emph {et~al.}(2010)\citenamefont
  {Grudinin}, \citenamefont {Lee}, \citenamefont {Painter},\ and\ \citenamefont
  {Vahala}}]{grudininPhononLaserAction2010}%
  \BibitemOpen
  \bibfield  {author} {\bibinfo {author} {\bibfnamefont {I.~S.}\ \bibnamefont
  {Grudinin}}, \bibinfo {author} {\bibfnamefont {H.}~\bibnamefont {Lee}},
  \bibinfo {author} {\bibfnamefont {O.}~\bibnamefont {Painter}},\ and\ \bibinfo
  {author} {\bibfnamefont {K.~J.}\ \bibnamefont {Vahala}},\ }\bibfield  {title}
  {\bibinfo {title} {Phonon laser action in a tunable two-level system},\
  }\href {https://doi.org/10.1103/PhysRevLett.104.083901} {\bibfield  {journal}
  {\bibinfo  {journal} {Phys. Rev. Lett.}\ }\textbf {\bibinfo {volume} {104}},\
  \bibinfo {pages} {083901} (\bibinfo {year} {2010})}\BibitemShut {NoStop}%
\bibitem [{\citenamefont {Zhang}\ \emph {et~al.}(2018)\citenamefont {Zhang},
  \citenamefont {Peng}, \citenamefont {{\"O}zdemir}, \citenamefont {Pichler},
  \citenamefont {Krimer}, \citenamefont {Zhao}, \citenamefont {Nori},
  \citenamefont {Liu}, \citenamefont {Rotter},\ and\ \citenamefont
  {Yang}}]{zhangPhononLaserOperating2018a}%
  \BibitemOpen
  \bibfield  {author} {\bibinfo {author} {\bibfnamefont {J.}~\bibnamefont
  {Zhang}}, \bibinfo {author} {\bibfnamefont {B.}~\bibnamefont {Peng}},
  \bibinfo {author} {\bibfnamefont {{\c S}.~K.}\ \bibnamefont {{\"O}zdemir}},
  \bibinfo {author} {\bibfnamefont {K.}~\bibnamefont {Pichler}}, \bibinfo
  {author} {\bibfnamefont {D.~O.}\ \bibnamefont {Krimer}}, \bibinfo {author}
  {\bibfnamefont {G.}~\bibnamefont {Zhao}}, \bibinfo {author} {\bibfnamefont
  {F.}~\bibnamefont {Nori}}, \bibinfo {author} {\bibfnamefont {Y.-x.}\
  \bibnamefont {Liu}}, \bibinfo {author} {\bibfnamefont {S.}~\bibnamefont
  {Rotter}},\ and\ \bibinfo {author} {\bibfnamefont {L.}~\bibnamefont {Yang}},\
  }\bibfield  {title} {\bibinfo {title} {A phonon laser operating at an
  exceptional point},\ }\href {https://doi.org/10.1038/s41566-018-0213-5}
  {\bibfield  {journal} {\bibinfo  {journal} {Nat. Photonics}\ }\textbf
  {\bibinfo {volume} {12}},\ \bibinfo {pages} {479} (\bibinfo {year}
  {2018})}\BibitemShut {NoStop}%
\bibitem [{\citenamefont {Zhang}\ \emph {et~al.}(2021)\citenamefont {Zhang},
  \citenamefont {Peng}, \citenamefont {Kim}, \citenamefont {Monifi},
  \citenamefont {Jiang}, \citenamefont {Li}, \citenamefont {Yu}, \citenamefont
  {Liu}, \citenamefont {Liu}, \citenamefont {Al{\`u}},\ and\ \citenamefont
  {Yang}}]{zhangOptomechanicalDissipativeSolitons2021}%
  \BibitemOpen
  \bibfield  {author} {\bibinfo {author} {\bibfnamefont {J.}~\bibnamefont
  {Zhang}}, \bibinfo {author} {\bibfnamefont {B.}~\bibnamefont {Peng}},
  \bibinfo {author} {\bibfnamefont {S.}~\bibnamefont {Kim}}, \bibinfo {author}
  {\bibfnamefont {F.}~\bibnamefont {Monifi}}, \bibinfo {author} {\bibfnamefont
  {X.}~\bibnamefont {Jiang}}, \bibinfo {author} {\bibfnamefont
  {Y.}~\bibnamefont {Li}}, \bibinfo {author} {\bibfnamefont {P.}~\bibnamefont
  {Yu}}, \bibinfo {author} {\bibfnamefont {L.}~\bibnamefont {Liu}}, \bibinfo
  {author} {\bibfnamefont {Y.-x.}\ \bibnamefont {Liu}}, \bibinfo {author}
  {\bibfnamefont {A.}~\bibnamefont {Al{\`u}}},\ and\ \bibinfo {author}
  {\bibfnamefont {L.}~\bibnamefont {Yang}},\ }\bibfield  {title} {\bibinfo
  {title} {Optomechanical dissipative solitons},\ }\href
  {https://doi.org/10.1038/s41586-021-04012-1} {\bibfield  {journal} {\bibinfo
  {journal} {Nature}\ }\textbf {\bibinfo {volume} {600}},\ \bibinfo {pages}
  {75} (\bibinfo {year} {2021})}\BibitemShut {NoStop}%
\bibitem [{\citenamefont {Carmon}\ \emph {et~al.}(2005)\citenamefont {Carmon},
  \citenamefont {Rokhsari}, \citenamefont {Yang}, \citenamefont {Kippenberg},\
  and\ \citenamefont
  {Vahala}}]{carmonTemporalBehaviorRadiationPressureInduced2005}%
  \BibitemOpen
  \bibfield  {author} {\bibinfo {author} {\bibfnamefont {T.}~\bibnamefont
  {Carmon}}, \bibinfo {author} {\bibfnamefont {H.}~\bibnamefont {Rokhsari}},
  \bibinfo {author} {\bibfnamefont {L.}~\bibnamefont {Yang}}, \bibinfo {author}
  {\bibfnamefont {T.~J.}\ \bibnamefont {Kippenberg}},\ and\ \bibinfo {author}
  {\bibfnamefont {K.~J.}\ \bibnamefont {Vahala}},\ }\bibfield  {title}
  {\bibinfo {title} {Temporal behavior of radiation-pressure-induced vibrations
  of an optical microcavity phonon mode},\ }\href
  {https://doi.org/10.1103/PhysRevLett.94.223902} {\bibfield  {journal}
  {\bibinfo  {journal} {Phys. Rev. Lett.}\ }\textbf {\bibinfo {volume} {94}},\
  \bibinfo {pages} {223902} (\bibinfo {year} {2005})}\BibitemShut {NoStop}%
\bibitem [{\citenamefont {L{\"u}}\ \emph {et~al.}(2015)\citenamefont {L{\"u}},
  \citenamefont {Jing}, \citenamefont {Ma},\ and\ \citenamefont
  {Wu}}]{luSymmetryBreakingChaosOptomechanics2015}%
  \BibitemOpen
  \bibfield  {author} {\bibinfo {author} {\bibfnamefont {X.-Y.}\ \bibnamefont
  {L{\"u}}}, \bibinfo {author} {\bibfnamefont {H.}~\bibnamefont {Jing}},
  \bibinfo {author} {\bibfnamefont {J.-Y.}\ \bibnamefont {Ma}},\ and\ \bibinfo
  {author} {\bibfnamefont {Y.}~\bibnamefont {Wu}},\ }\bibfield  {title}
  {\bibinfo {title} {{{PT}}-symmetry-breaking chaos in optomechanics},\ }\href
  {https://doi.org/10.1103/PhysRevLett.114.253601} {\bibfield  {journal}
  {\bibinfo  {journal} {Phys. Rev. Lett.}\ }\textbf {\bibinfo {volume} {114}},\
  \bibinfo {pages} {253601} (\bibinfo {year} {2015})}\BibitemShut {NoStop}%
\bibitem [{\citenamefont {Badzey}\ and\ \citenamefont
  {Mohanty}(2005)}]{badzeyCoherentSignalAmplification2005}%
  \BibitemOpen
  \bibfield  {author} {\bibinfo {author} {\bibfnamefont {R.~L.}\ \bibnamefont
  {Badzey}}\ and\ \bibinfo {author} {\bibfnamefont {P.}~\bibnamefont
  {Mohanty}},\ }\bibfield  {title} {\bibinfo {title} {Coherent signal
  amplification in bistable nanomechanical oscillators by stochastic
  resonance},\ }\href {https://doi.org/10.1038/nature04124} {\bibfield
  {journal} {\bibinfo  {journal} {Nature}\ }\textbf {\bibinfo {volume} {437}},\
  \bibinfo {pages} {995} (\bibinfo {year} {2005})}\BibitemShut {NoStop}%
\bibitem [{\citenamefont {Liu}\ and\ \citenamefont
  {Lai}(2001)}]{liuCoherenceResonanceCoupled2001}%
  \BibitemOpen
  \bibfield  {author} {\bibinfo {author} {\bibfnamefont {Z.}~\bibnamefont
  {Liu}}\ and\ \bibinfo {author} {\bibfnamefont {Y.-C.}\ \bibnamefont {Lai}},\
  }\bibfield  {title} {\bibinfo {title} {Coherence {{Resonance}} in {{Coupled
  Chaotic Oscillators}}},\ }\href {https://doi.org/10.1103/PhysRevLett.86.4737}
  {\bibfield  {journal} {\bibinfo  {journal} {Phys. Rev. Lett.}\ }\textbf
  {\bibinfo {volume} {86}},\ \bibinfo {pages} {4737} (\bibinfo {year}
  {2001})}\BibitemShut {NoStop}%
\bibitem [{\citenamefont {Li}(2024)}]{liPositiveIncentiveNoise2024}%
  \BibitemOpen
  \bibfield  {author} {\bibinfo {author} {\bibfnamefont {X.}~\bibnamefont
  {Li}},\ }\bibfield  {title} {\bibinfo {title} {Positive-{{Incentive
  Noise}}},\ }\href {https://doi.org/10.1109/TNNLS.2022.3224577} {\bibfield
  {journal} {\bibinfo  {journal} {IEEE Trans. Neural Netw. Learn. Syst.}\
  }\textbf {\bibinfo {volume} {35}},\ \bibinfo {pages} {8708} (\bibinfo {year}
  {2024})}\BibitemShut {NoStop}%
\bibitem [{\citenamefont {Shi}\ \emph {et~al.}(2025)\citenamefont {Shi},
  \citenamefont {Lv}, \citenamefont {Fu}, \citenamefont {Wang}, \citenamefont
  {Huang}, \citenamefont {Wei}, \citenamefont {Amabili},\ and\ \citenamefont
  {Huan}}]{shiNoiseenhancedStabilitySynchronized2025}%
  \BibitemOpen
  \bibfield  {author} {\bibinfo {author} {\bibfnamefont {Z.}~\bibnamefont
  {Shi}}, \bibinfo {author} {\bibfnamefont {Q.}~\bibnamefont {Lv}}, \bibinfo
  {author} {\bibfnamefont {M.}~\bibnamefont {Fu}}, \bibinfo {author}
  {\bibfnamefont {X.}~\bibnamefont {Wang}}, \bibinfo {author} {\bibfnamefont
  {Z.}~\bibnamefont {Huang}}, \bibinfo {author} {\bibfnamefont
  {X.}~\bibnamefont {Wei}}, \bibinfo {author} {\bibfnamefont {M.}~\bibnamefont
  {Amabili}},\ and\ \bibinfo {author} {\bibfnamefont {R.}~\bibnamefont
  {Huan}},\ }\bibfield  {title} {\bibinfo {title} {Noise-enhanced stability in
  synchronized systems},\ }\href {https://doi.org/10.1126/sciadv.adx1338}
  {\bibfield  {journal} {\bibinfo  {journal} {Sci. Adv.}\ }\textbf {\bibinfo
  {volume} {11}},\ \bibinfo {pages} {eadx1338} (\bibinfo {year}
  {2025})}\BibitemShut {NoStop}%
\bibitem [{\citenamefont {Li}\ \emph {et~al.}(2023)\citenamefont {Li},
  \citenamefont {Li}, \citenamefont {Xiong}, \citenamefont {Xu}, \citenamefont
  {Wang}, \citenamefont {Tian}, \citenamefont {Yang}, \citenamefont {Liu},
  \citenamefont {Zeng}, \citenamefont {Lin}, \citenamefont {Li}, \citenamefont
  {Lee}, \citenamefont {Ho},\ and\ \citenamefont
  {Qiu}}]{liStochasticExceptionalPoints2023}%
  \BibitemOpen
  \bibfield  {author} {\bibinfo {author} {\bibfnamefont {Z.}~\bibnamefont
  {Li}}, \bibinfo {author} {\bibfnamefont {C.}~\bibnamefont {Li}}, \bibinfo
  {author} {\bibfnamefont {Z.}~\bibnamefont {Xiong}}, \bibinfo {author}
  {\bibfnamefont {G.}~\bibnamefont {Xu}}, \bibinfo {author} {\bibfnamefont
  {Y.~R.}\ \bibnamefont {Wang}}, \bibinfo {author} {\bibfnamefont
  {X.}~\bibnamefont {Tian}}, \bibinfo {author} {\bibfnamefont {X.}~\bibnamefont
  {Yang}}, \bibinfo {author} {\bibfnamefont {Z.}~\bibnamefont {Liu}}, \bibinfo
  {author} {\bibfnamefont {Q.}~\bibnamefont {Zeng}}, \bibinfo {author}
  {\bibfnamefont {R.}~\bibnamefont {Lin}}, \bibinfo {author} {\bibfnamefont
  {Y.}~\bibnamefont {Li}}, \bibinfo {author} {\bibfnamefont {J.~K.~W.}\
  \bibnamefont {Lee}}, \bibinfo {author} {\bibfnamefont {J.~S.}\ \bibnamefont
  {Ho}},\ and\ \bibinfo {author} {\bibfnamefont {C.-W.}\ \bibnamefont {Qiu}},\
  }\bibfield  {title} {\bibinfo {title} {Stochastic {{Exceptional Points}} for
  {{Noise-Assisted Sensing}}},\ }\href
  {https://doi.org/10.1103/PhysRevLett.130.227201} {\bibfield  {journal}
  {\bibinfo  {journal} {Phys. Rev. Lett.}\ }\textbf {\bibinfo {volume} {130}},\
  \bibinfo {pages} {227201} (\bibinfo {year} {2023})}\BibitemShut {NoStop}%
\end{thebibliography}%

\end{document}